\documentclass[12pt]{iopart}
\usepackage{iopams}
\usepackage{graphicx}
\usepackage{bm}
\usepackage{iopams,setstack}
\usepackage{mathrsfs}
\usepackage[usenames]{color}
\definecolor{quique}{rgb}{0.1,0,0.95}

\newcommand{\lag}{\ensuremath{\mathscr{L}}}        

\newcommand{\D}{\displaystyle}
\newcommand{\beq}{\begin{equation}}
\newcommand{\eeq}{\end{equation}}
\newcommand{\beqn}{\begin{equation*}}
\newcommand{\eeqn}{\end{equation*}}
\newcommand{\beqa}{\begin{eqnarray}}
\newcommand{\eeqa}{\end{eqnarray}}
\newcommand{\beqan}{\begin{eqnarray*}}
\newcommand{\eeqan}{\end{eqnarray*}}
\def\nn{\nonumber\\}
\def\eq#1{(\ref{#1})}
\def\cuad{\hfill $\Box$\\}
\def\cd#1{\ensuremath{\nabla_{#1}}}          
\def\pd#1{\ensuremath{\partial_{#1}}}        

\def\dlag#1{\frac{\partial\lag}{\partial\ \! #1}}

\def\st{spacetime}

\def\mch{\scriptscriptstyle}

\def\text#1{{\rm #1}}

\def\lab#1{\label{#1}}

\def\A{\mathcal{A}}
\def\vs#1{\vspace{#1 cm}}
\def\hs#1{\hspace{#1 cm}}

\newtheorem{theorem}{Theorem}[section]
\newtheorem{lem}[theorem]{Lemma}
\newtheorem{definition}[theorem]{Definition}

\newtheorem{cor}[theorem]{Corollary}
\newtheorem{rem}[theorem]{Remark}
\newtheorem{notation}[theorem]{Notation}

\def\b{\mathfrak{b}}

\def\R{\mathbb{R}}
\def\O{\Omega}

\def\E{\mathcal{E}}

\def\M{\mathscr{M}}
\def\dis{\displaystyle}
\def\N{\mathbb{N}}

\begin{document}

\title[On well-posedness of the Cauchy problem \dots]{On  well-posedness of the Cauchy problem for the wave equation in static
spherically symmetric spacetimes}

\author{Ricardo\ E  Gamboa Sarav\'{\i}$^{1,2}$,  Marcela Sanmartino$^3$ and  Philippe Tchamitchian $ ^4$}
\address{$^1$ Departamento de F\'{\i}sica, Facultad de Ciencias
Exactas, Universidad Nacional de La Plata, Casilla de Correo 67, 1900 La Plata, Argentina. }
\address{$^2$ IFLP, CONICET, La Plata,
Argentina.  }
\address{$^3$ Departamento de Matem\'{a}tica, Facultad de Ciencias
Exactas, Universidad Nacional de La Plata, Casilla de Correo 172, 1900 La Plata, Argentina. }
\address{$^4$ Aix-Marseille Universit\'e, CNRS, LATP (UMR 6632), 39, rue F. Joliot-Curie,
13453 Marseille Cedex 13, France}

\ead{quique@fisica.unlp.edu.ar, tatu@mate.unlp.edu.ar, philippe.tchamitchian@univ-amu.fr}

\date{\today}

\begin{abstract}
We give simple conditions implying the well-posedness of the Cauchy problem for the propagation of classical scalar fields in general $(n+2)$-dimensional static and spherically symmetric spacetimes. They are related to properties of the underlying spatial part of the wave operator, one of which being the standard essentially self-adjointness. However, in many examples  the spatial part of the wave operator turns out to be not essentially selfadjoint, but it does satisfy a weaker property that we call here {\it quasi essentially self-adjointness}, which is enough to ensure the desired well-posedness. This is why we also characterize this second property.

We state abstract results, then general results for a class of operators encompassing many examples in the literature, and we finish with the explicit analysis of some of them.
\end{abstract}


\section{Introduction}

Hawking and Penrose have shown that, according to general relativity, there must exist singularities of infinite density and space-time curvature in many physically reasonable situations.  This phenomenon  occurs in the big bang scenery at the very beginning of time, and it would be an end of time for sufficiently massive collapsing bodies (see, for example, \cite{Haw} and references therein). At these singularities all the known laws of physics and our ability to predict the future would break down.

However, in the case of black holes, any observer who remained outside the event horizon would not be affected by this failure of predictability, because neither light nor any other signal could reach him from the singularity. This  notable feature led Penrose to propose the weak {\em  cosmic censorship hypothesis}: all singularities produced by gravitational collapse occur only in places, like black holes, where they are hidden from outside view by an event horizon \cite{P1}.

The strong version of the cosmic censorship hypothesis states that any physically  realistic spacetime must be globally hyperbolic \cite{P2}. The concept of global hyperbolicity was introduced for dealing with hyperbolic partial  differential equations on a manifold \cite{Le}. A spacetime is said to be globally hyperbolic if, given any two of its points, the set of of all causal curves joining these points is compact (in a suitable topology). Only in this case there is a Cauchy surface whose domain of dependence is the entire spacetime. This is a reasonable condition to impose, for example, to ensure the existence and uniqueness of solutions of hyperbolic differential equations \cite{Le,CB}.

Nevertheless, the relevant physical condition to assure predictability is not global hyperbolicity, but the well-posedness of the field equations. Indeed, there are  many examples  of spacetimes that are not geodesically complete and violate cosmic censorship, but where there is still a well-posed initial-value
problem for test fields. Global hyperbolicity is sufficient, but not necessary for this. This suggests that, in more general situations, we could  find a weaker condition to replace the notion of global hyperbolicity  by making direct reference to test fields \cite{W,HM,Cl}.

The above considerations motivate a  deeper study of the well-posedness of the initial-value problem for fields in more general singular spacetimes.

This paper is a continuation of a previous one \cite{GST}, tackling the well-posedness of Cauchy problem for waves in static spacetimes.
This subject has been launched by Wald in \cite{W}, and
further developed by, among others, the authors of references
\cite{HM, S,SST}.

The propagation of waves is, in such spaces, described by a classical equation of the form
$$\partial_{tt} \phi + \A \phi = 0,$$
where $\A$ is a selfadjoint extension of a given symmetric and positive operator $A$ which reflects the underlying geometry.

Our motivation relies on the following observation: although $A$ may not be essentially selfadjoint ({\it e.s.a.}), boundary conditions are not necessary to construct $\A$ in some geometries of interest. Such a situation  arises when, even if $A$ has many selfadjoint extensions, only one has its domain included in the energy space naturally associated to $A$. Here we call {\it quasi essentially selfadjoint} ({\it q.e.s.a.}) this property.

We have shown in \cite{GST} that operators $A$ given by propagation of massless scalar fields in static spacetimes with naked timelike singularities may be  {\it q.e.s.a.} but not {\it e.s.a.}. Thus, in such situations, demanding the finiteness of the energy is enough to select one selfadjoint extension of $A$, and only one; in addition, we proved that the solutions of the wave equation may have a non trivial trace at the boundary of the geometrical domain, even though this trace is not imposed by any boundary condition at all. This phenomenon never happens with {\it e.s.a.} operators.

Here we deeply examine the case of general $(n+2)$-dimensional static and spherically symmetric spacetimes. More precisely, the concrete setting is the following.

The domain is of the form  $I \times  \M$, where $I \subset (0,+\infty)$ is an open interval and $\M$ is a compact, oriented  Riemannian manifold without boundary. The operator $A$ is defined on $C_0^{\infty}(I \times \M)$ as

\beq\fl\lab{ope}
A\varphi(z, \mathbf{x})=\frac{1}{a(z)}\left\{- \pd z \Bigl(b(z) \pd z \varphi(z, \mathbf{x})\Bigr)- c(z) \Delta_{\hs {-.07}\mch \M} \hs {.07}\varphi(z, \mathbf{x})+ d(z)\varphi(z, \mathbf{x})\right\},
\eeq
where $\Delta_{\hs {-.07}\mch \M} \hs {.07}$ is the Laplace-Beltrami operator on $\M$, and $a$, $b$, $c$ and $d$ are suitable positive coefficients only depending on the radial variable $z \in I$. No condition is prescribed on the coefficients at the boundary of the domain.

For this class of operators we fully characterize {\it e.s.a.} and {\it q.e.s.a.} properties. More precisely,  under rather general conditions on the coefficients, we give a necessary and sufficient condition for {\it q.e.s.a.} depending only on the integrability of the function $\displaystyle\left(\frac1{b(z)}+d(z)+a(z)\right)$ at the boundary of $I$. We also give a necessary and sufficient condition for {\it e.s.a.}, in this case the condition depends also on the integrability of the functions $a(z)$ and  $\beta(z)^2 a(z)$ at the boundary of $I$, where  $\beta(z)$ is a particular solution of the ordinary differential equation $- \Bigl(b(z)\,\beta'(z)\Bigr)'+d(z)\, \beta(z)=0$.

We then apply this analysis to scalar fields propagating in static spherically symmetric spacetimes of arbitrary dimension, solutions of the Einstein equations with cosmological constant and  matter satisfying the dominant energy condition or vacuum. The criteria for {\it e.s.a.} and {\it q.e.s.a.}  on the coefficients of the operator $A$ are then translated into criteria on the components of the metric tensor.
This  provides a systematic procedure to analyze the situations where boundary conditions are, or are not, necessary for the Cauchy problem to be well-posed.

A significant physical result is stated in theorem 5.5: in the outer region of a static, spherically symmetric and asymptotically flat spacetime where the dominant energy condition holds, the operator $A$ is essentially selfadjoint, i.e. the Cauchy problem is well-posed without any boundary conditions, if, and only if, an observer at infinity measures that it takes an infinite time to a photon to reach the boundary.

Finally, we directly apply the developed theory  to the discussion of some exact vacuum solutions as explicit examples. We discuss the $(n+2)$-dimensional  Minkowski \st\ with a removed spatial point and the higher-dimensional generalization of Schwarzschild and Reissner-Nordstr\"{o}m geometries; we systematically describe the situations where boundary conditions are, or are not, necessary for the Cauchy problem to be well-posed.

The outline of the paper is as follows. Section 2 is devoted to abstract results on {\it e.s.a.} and {\it q.e.s.a.} properties. In section 3 we completely characterize {\it e.s.a.} and {\it q.e.s.a.} properties of the operator given in \eq{ope}. We show,  in section 4, the well-posedness of the Cauchy problem when the operator $A$ is {\it q.e.s.a.}  but not necessarily {\it e.s.a.}. In section 5 we  apply our results to the study of propagation of scalar fields in general $(n+2)$-dimensional static and spherically symmetric \st\ with $n\geq 1$. We close by discussing  the examples in section 6.

\section{Quasi essentially  and essentially selfadjointness}\lab{S2}

Let $\Omega\subset \R^{n+1}$  be a Lipschitz domain\footnote{Being Lipschitz is not the weakest possible hypothesis on $\Omega$ for our results to hold, but it is enough for the examples we have in mind.} and $H$ a Hilbert space such that $C_c^{\infty}(\Omega)$ is dense in $H$,  where $C_c^{\infty}(\Omega)$ is the space of the restrictions to $\Omega$ of $C_0^{\infty}(\R^{n+1})$. We consider an unbounded symmetric definite positive operator $A$, whose  domain is $C_0^{\infty}(\Omega)$.
We assume the existence of  a Hilbert space $\E$, continuously embedded in $H$, and a related bilinear symmetric form $\b$  with domain $\E$ having the
following properties:
\begin{enumerate}
\item if $\phi\in\E$, $\|\phi\|_{\E}^2=\|\phi\|^2_H+\b(\phi,\phi)$;
\item $C_c^\infty(\Omega)$ is dense in $\E$;
\item if $\phi, \psi \in C_0^\infty(\Omega)$, then $ \b(\phi, \psi)=\langle\phi,A\psi\rangle$.
\end{enumerate}

The reader should note that $A$ is defined only on $C_0^\infty (\Omega)$, and that consequently
the relation between the form $\b$ and the operator $A$ is only stated for functions
in $C_0^\infty (\Omega)$ as well, although $C_c^\infty (\Omega)$ is dense in both spaces $H$ and $\E$. This is motivated by the difficulties arising with boundary conditions: whether they must
be specified in advance or not is the question we consider in the subsequent
theorem 2.2. We will show that there is a ``natural" self-adjoint extension of $A$,
defined without specifying any boundary condition, if and only if $C_0^\infty (\Omega)$
is dense in $\E$. We will also show that this density property is always true when $A$
is essentially self-adjoint, but may occur even when $A$ is not. Various examples
are given at the end of the paper.
\begin{definition}
We shall  say that $\A$, any given selfadjoint
extension of $A$, is of finite energy when $D(\A) \subset \E$, with continuous injection.
\end{definition}

 Calling  $\E^0$ the closure of $C_0^\infty(\Omega)$ in $\E$, we have the following result:

 \begin{theorem}\label{general}
 Under these hypotheses we have:
\begin{enumerate}
\item The operator $A$ has only one selfadjoint extension  with finite energy if
and only if $\E^0 = \E$. If this is the case, this extension is $A_{\mch F}$, the Friedrichs extension.

\item If $\E^0 = \E$, then  $C_0^{\infty}(\Omega)$ is dense in $D(A_{\mch F})$ if and only if $A$ is essentially selfadjoint (e.s.a.), i.e., $A$ has only one selfadjoint extension.

\end{enumerate}

\end{theorem}

\noindent{\bf Proof:}

\textit{(i)}  To prove this assertion, we begin with assuming that $A$
has only one selfadjoint extension with finite energy. Let $\mathcal{A}$ be the selfadjoint operator associated with the energy
form $\b$; let
$\mathcal{A}_0$ be the selfadjoint operator associated with the restriction of $\b$ to  $\E^0$.
Both are extensions of $A$ with domains included in $\E$, and so, are equal.
But then we must have $D(\mathcal{A}^\frac12)=D(\mathcal{A}_0^\frac12)$, which is $\E = \E^0$.

Reciprocally, if $\E = \E^0$, the only selfadjoint
extension of $A$ with domain in $\E$ is its Friedrichs extension,
because the form $\b$ defined on $\E$ is the closure of the form
$\b$ defined on $C_0^{\infty}(\Omega)$.
\smallskip

\textit{(ii)}
  Recall that
\begin{equation*}\label{dominio*}
D(A^*)=\{\varphi \in H  : \exists\, C> 0\  : \forall  \psi \in C_0^{\infty}(\O),\ |\langle\varphi, A\psi\rangle|\leq C \|\psi\|_H\},
\end{equation*}
and that
\begin{equation}\label{dominioF}
D(A_F)=\{\varphi \in \E : \exists\, C> 0\ : \forall \eta \in \E ,\ \mathfrak{b}(\varphi, \eta)\leq C \|\eta\|_H\}.
\end{equation}We assume first that $C_0^{\infty}(\Omega)$ is dense in $D(A_{\mch F})$. It is enough to see that $D(A^*)\subset D(A_{\mch F})$. Taking $\phi_0\in D(A^*)$ and $\eta_0=(A^*+ I)\phi_0$, we have for all $\psi \in  C_0^{\infty}(\Omega)$
$$\langle\phi_0, (A_{\mch F}+I)\psi\rangle= \langle\phi_0, (A+I)\psi\rangle=\langle\eta_0, \psi\rangle$$ and then, since $C_0^{\infty}(\Omega)$ is dense in  $D(A_{\mch F})$,  for all $\varphi \in D(A_{\mch F})$
$$ \langle\phi_0, (A_{\mch F}+I)\varphi\rangle=\langle\eta_0, \varphi\rangle.$$

Taking into account that $(A_{\mch F}+I)^{-1}$  is defined on all $H$, by calling $\varphi_0=(A_{\mch F}+I)^{-1}\eta_0 \in D(A_{\mch F})$   we have
$$\langle\eta_0, \varphi\rangle= \bigl\langle(A_{\mch F}+I)(A_{\mch F}+I)^{-1}\eta_0,\varphi\rangle=\langle\varphi_0, (A_{\mch F}+I)\varphi\rangle  \ \ \mbox{for all}\ \varphi \in D(A_{\mch F}),$$
and then
$$\langle\varphi_0-\phi_0, (A_{\mch F}+I)\varphi\rangle=0\ \ \ \mbox{for all} \ \varphi \in D(A_{\mch F}).$$

Since $\mbox{Im}(A_{\mch F}+I)=H$, we have $\varphi_0=\phi_0$. It implies $D(A^*)\subset D(A_{\mch F})$ and so $A^*=A_{\mch F}$. Then $A$ is essentially selfadjoint.
\ \\

On the other hand, if  $C_0^{\infty}(\Omega)$ is not dense in $D(A_{\mch F})$, there exists $\varphi \in D(A_{\mch F})$ such that $A_{\mch F}\varphi\neq 0$ and $$\langle A_{\mch F}\varphi,A_{\mch F}\psi\rangle=0\ \ \forall \psi \in C_0^{\infty}(\Omega).$$
  Let us call $\eta=A_{\mch F}\varphi$. If $\eta \in \E$, then $\b(\eta,\psi)=\langle\eta,A\psi\rangle=\langle\eta,A_{\mch F}\psi\rangle=0$ for all $\psi \in C_0^{\infty}(\Omega)$ and then by density of $C_0^{\infty}(\Omega)$ in $\E$, $\b(\eta,\eta)=0$. Since by hypothesis $\eta\neq 0$, we have $\eta \notin \E$.

Therefore, we have proved that there exists $\eta\in H$, such that $\eta\in \ker(A^*)$ but $\eta \notin \E$, so $A$ cannot be essentially self adjoint.\cuad

\begin{definition} Under the preceding hypotheses, the operator $A$ is \textbf{quasi essentially selfadjoint} (q.e.s.a.) if it has only one extension with finite energy.
\end{definition}

\begin{lem}\label{dominio}
If $A$ is a {\it q.e.s.a.} operator, then $\displaystyle{D(A_F)=D(A^*)\cap \E}$.
\end{lem}
{\bf Proof:}

Since $\displaystyle{D(A_F)\subset D(A^*)}$ by definition of $A^*$ and $\displaystyle{D(A_F) \subset \E}$ by definition of $A_F$, then $\displaystyle{D(A_F)\subset D(A^*)\cap \E}$.

Conversely, let $\varphi \in D(A^*)\cap \E$, then
\begin{equation*}
\mathfrak{b}(\varphi, \psi)\leq C \|\psi\|_H \ \ \ \forall \psi \in C_0^{\infty}(\O)
\end{equation*}
by definition of $D(A^*)$. Since $C_0^{\infty}(\O)$ is dense in $\E$ and $\varphi \in \E$, this inequality extends to any $\psi \in \E$, proving that $\varphi \in D(A_F)$.\cuad

\begin{lem}\label{equivalente} If $A$ is a  {\it q.e.s.a.} operator, then the three following statements are equivalent

\begin{enumerate}
\item $A$ is not an e.s.a. operator.
\item there exists $\varphi \in D(A^*)$ but $\varphi \notin \E$.
\item there exists $\varphi \in D(A^*)$  non vanishing and such that $\displaystyle{(A^*+I)\varphi=0}$.
\end{enumerate}
\end{lem}

\noindent{\bf Proof:}

{\it (i)} $\Leftrightarrow$ {\it (ii)}:
Observe that  $A$ is an {\it e.s.a.} operator if and only if  $A^*=A_F$, thus, by lemma \ref{dominio}, $A$ is {\it e.s.a.} operator if and only if $D(A^*)\subset \E$.

 {\it (ii)} $\Leftrightarrow$ {\it (iii)}
 Let $\varphi_0 \in D(A^*)$ and $\varphi_0\notin \E$, and define $f=(A^*+I)\varphi_0 \in H$, $\varphi=(A_{\mch F}+I)^{-1}f\in D(A_{\mch F} )$. We have
$
(A_{\mch F}+I)\varphi=(A^*+I)\varphi_0\,
$
and since $\varphi \in D(A^*)$, this implies $A^*(\varphi_0-\varphi) + (\varphi_0-\varphi)=0$. Finally $\varphi_0-\varphi$ cannot identically vanish, since $\varphi_0\notin \E$ while $\varphi \in \E$. Thus {\it (iii)} holds.

Conversely, let $\varphi\neq 0$ a.e., $\varphi \in D(A^*)$ such that $(A^*+I)\varphi=0$. If  $\varphi \in \E$, by {lemma~\ref{dominio}},  $\varphi \in D(A_{\mch F} )$ , then $\varphi = 0$ a.e. since  $A_{\mch F}+I$ is injective, which is a contradiction.  Thus, $\varphi \notin \E$ and  {\it (ii)} holds.\cuad

\section{A characterization of  some {\it  q.e.s.a.} and {\it  e.s.a.} divergence type operators}\lab{S3}

Let $\M$ be  a  Riemannian manifold of dimension $n$ with a metric $(g_{ij})$. We  also assume that    $\M$ is compact, connected, without boundary and with a given orientation.

In local coordinates, for $u\in C^{\infty}(\M)$ the Laplace-Beltrami operator is
\begin{equation*}
\Delta_{\hs {-.07}\mch \M} u={\rm div}(\nabla_{\hs {-.15}\mch \M}\hs {.05}u)=\frac{\sum_{i,j=1}^n\pd i \left(\sqrt{g}\,g^{ij}\pd j u \right)}{\sqrt{g},}\,,
\end{equation*}
where $g$ is the determinant of the metric. Let us consider in  $\Omega=(0,+\infty)\times  \M$,
 the operator $A$ given by
\beq\label{operador}\fl
A\varphi(z, \mathbf{x})=\frac{1}{a(z)}\left\{- \pd z \Bigl(b(z) \pd z \varphi(z, \mathbf{x})\Bigr)- c(z) \Delta_{\hs {-.07}\mch \M} \hs {.07}\varphi(z, \mathbf{x})+ d(z)\varphi(z, \mathbf{x})\right\},
\eeq
 for all $\varphi \in C_0^{\infty}(\Omega)$, where the functions $a, b, c$ and $d$ satisfy the following hypotheses:
\begin{itemize}\label{hipotesis}
 \item $a$ , $c$ , $d \in  L_{loc}^1\left(\rule{0pt}{10.5pt}(0,+\infty)\right)$ and $ b \in  C\left(\rule{0pt}{10.5pt}(0,+\infty)\right)$,
 \item $a>0,\, b>0,\, c>0\, \mbox{and} \, d\geq 0$ in $(0,+\infty)$,
 \item $a^{-1},\, b^{-1},\, c^{-1} \in L_{loc}^1\left(\rule{0pt}{10.5pt}(0,+\infty)\right)$.
\end{itemize}

Examples will be presented in the two last sections. Let us state in advance that the coefficient $d$ is non vanishing only in the massive case. This is why we will call {\it massless} the case $d=0$.

We define the Hilbert spaces
 $$
 H=\{\varphi \in L^2_{loc}(\O): \int_{\O}|\varphi(z, \mathbf{x})|^2 a(z) d\omega_{\hs {-.07}\mch \M}\,dz <\infty \},
 $$
and the energy space
 $$
 \E = \{\varphi \in H \cap H^1_{loc}(\O): \b(\varphi,\varphi)<+\infty \},
 $$
where we denote $\omega_{\hs {-.08}\mch \M}$  the natural measure in  $\M$, and
\beqan\label{bL}\fl
\b(\varphi,\psi)= \int_{\O} b(z)\,\pd z \varphi(z, \mathbf{x})\, \overline{\pd z \psi (z, \mathbf{x})}\, d\omega_{\hs {-.08}\mch \M}\, dz\, + \int_{\O} c(z)\,\nabla_{\hs {-.15}\mch \M}\,\varphi(z, \mathbf{x})
\cdot \overline{\nabla_{\hs {-.15}\mch \M}\psi(z, \mathbf{x})} \, d\omega_{\hs {-.08}\mch \M}\,dz\nn \hs {-1.2}+\int_{\O} d(z)\,\varphi(z, \mathbf{x})
\, \overline{\psi(z, \mathbf{x})} \, d\omega_{\hs {-.08}\mch \M}\,dz \,,
\eeqan
for $\varphi, \psi\in C_0^{\infty}(\O)$.

Thus, $H$ and $\E$ are  Hilbert spaces, equipped with their canonical norms:
$\displaystyle{\|\varphi\|_H ^2= \int_{\O}|\varphi(z, \mathbf{x})|^2 a(z)\, d\omega_{\hs {-.08}\mch \M}\,dz }$ and  $\displaystyle{\|\varphi\|_{\E}^2= \|\varphi\|^2_H+ \b(\varphi, \varphi)}$. The operator $A$ is well defined on $C_0^{\infty}(\O)$ and it is symmetric in $H$ by definition.

We shall explore when $A$ is a {\it q.e.s.a.} operator by using Theorem \ref{general}. Then the question is to determine under which conditions on the coefficients of $A$, $C_0^{\infty}(\O)$  is dense in $\E$. A related one is whether  $C_c^{\infty}(\overline{\O})\cap \E$ is dense in $\E$.

\begin{notation}\rm
From now on, $\dis \int_{z_0} $ and  $\dis  \int^{z_1}$  respectively denote $\dis  \int_{z_0}^{z_0+\varepsilon} $ and $\dis  \int_{z_1-\varepsilon}^{z_1} $   for a positive and  small enough $\varepsilon$. And $\D{\int^{+\infty} < + \infty}$
means that their exists $z>0$ such that
$\D{\int_z^{+\infty} < + \infty}$.

\end{notation}

\begin{theorem}\label{q.e.s.a.} Let $A$ be the operator defined in \eq{operador}. Then
\begin{enumerate}
\item $C_c^{\infty}(\overline{\O})\cap \E$ is dense in $\E$ if and only if $\displaystyle{\int^{+\infty}\left(\frac1{b(z)}+d(z)+a(z)\right)dz=+\infty}$,
\item $A$ is a q.e.s.a. operator ({\it i.e.} $C_0^{\infty}(\O)$ is dense in $\E$) if and only if $\displaystyle{\int^{+\infty}\left(\frac1{b(z)}+d(z)+a(z)\right)dz=+\infty}$ and $\displaystyle{\int_0\left(\frac1{b(z)}+d(z)+a(z)\right)dz=+\infty}$.
\end{enumerate}

\end{theorem}

\noindent{\bf Proof:}

The proof goes through three steps: first reducing the problem to a one dimensional case, second proving that compactly supported functions are dense under the given hypotheses, and finally getting the desired result.
\subsection*{\bf Step 1: reduction to the one dimensional case.}

Let $\{\lambda_k,\, k\geq 0 \}$ be the spectrum of $-\Delta_{\hs {-.07}\mch \M}$, with $\lambda_0=0$ and $\lambda_k$ an increasing sequence, and let $(\psi_k)_{k\geq 0}$ be an associated orthonormal basis of $L^2(\M)$.

 We define, for each $k\geq 0$,
\beq\label{Ak}
 A_k\, u(z)=\frac1{a(z)}\Biggl( -\biggl(b(z)\,u'(z)\Bigr)'+\Bigl(\lambda_k\,{c(z)}+{d(z)}\Bigr)\,u(z)\Biggr)\,,
 \eeq
 for $u\in C_0^{\infty}\left(\rule{0pt}{10.5pt}(0,+\infty)\right)$, with the underlying Hilbert space $H_0=L^2\left(\rule{0pt}{10.5pt}(0,+\infty),a(z)\,dz\right)$ and energy spaces $\E_k= \left\{u \in H_0 \cap H^1_{loc}\left(\rule{0pt}{10.5pt}(0,+\infty)\right): \b_k(u,u)<+\infty\right \}$, where
\beqn\fl{\mathfrak{b}_k(u,v)}=\int_0^{+\infty}b(z)\,u'(z)\,\overline{v'(z)}\,dz+\int_0^{+\infty}\Bigl(\lambda_k\,c(z)+d(z)\Bigr)\,u(z)\,\overline{v(z)}\,dz\,.\eeqn Then we consider the Hilbert spaces $\E_k$ with their natural norms
\begin{equation*}
\|u\|^2_{\E_k}=\int_0^{+\infty}b(z)\,|u'(z)|^2\,dz+\int_0^{+\infty}\Bigl(\lambda_k\,c(z)+d(z)+a(z)\Bigr)\,|u(z)|^2\,dz.
\end{equation*}

\begin{lem}\label{Ek}
$\displaystyle{C_c^{\infty}(\overline{\O})\cap \E}$ (respectively $C_0^{\infty}(\O)$) is dense in $\E$   if and only if $C_c^{\infty}\left(\rule{0pt}{10.5pt}[0,\infty)\right)\cap \E_k$ (respectively $C_0^{\infty}\left(\rule{0pt}{10.5pt}(0,+\infty)\right)$ is dense in $\E_k$  for all $k \geq 0$.
\end{lem}

\noindent{\bf Proof:}

Given $\varphi\in \E$, it can be decomposed into a sum $\varphi= \displaystyle{\sum_{k\geq 0}u_k\otimes\psi_k}$, where $u_k\in \E_k$ and

\beqn
\|\varphi\|^2_\E= \sum_{k\geq 0}\|u_k\|_{\E_k}^2.
\eeqn
So, density in $\E$ implies density in each $\E_k$.

For the reciprocal, given $\varphi \in \E$ we first approximate it by the functions $\varphi_m= \displaystyle{\sum_{k=0}^mu_k\otimes\psi_k}$, and density in $\E_k$ for all $k\geq0$ implies  that each $\varphi_m$ can be approximate by functions of $\displaystyle{C_c^{\infty}(\O)\cap \E}$ (respectively $C_0^{\infty}(\O)$).\cuad

\subsection*{\bf Step 2: density of compactly supported functions in $\E_0$.}
Here, for convenience  we shall restrict our attention at first to the case $k=0$ and $d(z)\equiv0$.

We define
\beqan
 \E_{0,c}=\E_0\cap\{\mbox{functions with compact support in } [\,0,+\infty)\}, \\
 \E_{0,0}=\E_0\cap\{\mbox{functions with compact support in } (0,+\infty)\}.
\eeqan

\begin{lem}\label{densidad1}
$ \E_{0,c}$ is dense in $\E_0$ if and only if $\dis{\int^{+\infty}\left(\frac1{b(z)}+a(z)\right)\,dz=+\infty}$.
\end{lem}
\noindent{\bf Proof:}

Assume first that $\dis{\int^{+\infty}\left(\frac1{b(z)}+a(z)\right)\,dz<+\infty}$. If $u \in \E_0$, \mbox{then $u'\in L^1\left(\rule{0pt}{10.5pt}\,[z',+\infty)\right)$} for any $z'>0$, since $\dis{\int_{z'}^{+\infty}\frac1{b(z)}\,dz<+\infty}$ and using H\"{o}lder inequality. Moreover $\displaystyle{\lim_{z\to \infty}u(z)}$ exists and is not necessarily zero because $\dis\int^{+\infty}a(z)<+\infty$. Thus, there exists a linear functional on $\E_0$ which vanishes on $\E_{0,c}$ but not everywhere, showing that $ \E_{0,c}$ is not dense in $\E_0$.
Such functional may be
\beqn
\lambda(u)=\int_0^{+\infty}\Bigl(u(z)\,\eta(z)\Bigr)' dz,
\eeqn
where $\eta(z)$ is a smooth function such that $\eta(z)=0$ if  $z\in [0, z']$ and $\eta(z)=1$ if $z\geq 2z'$.

Assuming now that $\dis{\int^{+\infty}\left(\frac1{b(z)}+a(z)\right)\,dz=+\infty}$,  we shall see that $ \E_{0,c}$ is dense in $\E_0$.

If there exists $z'>0$ such that $\displaystyle{\int_{z'}^{+\infty}\frac1{b(z)}\,dz<+\infty}$, taking $u\in \E_0$, we have again that $u'\in L^1\left(\rule{0pt}{10.5pt}\,[z',+\infty)\right)$, but now $\dis \lim_{z\to +\infty}u(z)=0$ necessarily, since $\dis{\int^{+\infty}a(z)\,dz=+\infty}$. Thus, we have
\beqn
u(z)=-\int_z^{+\infty}u'(s)\,ds.
\eeqn
Hence, defining $\displaystyle{\beta_0(z)=\int_z^{+\infty}\frac1{b(z)}\,dz}$ and using H\"{o}lder inequality we have
\beqa\label{acotphi}
|u(z)| \leq   \sqrt{\beta_0(z)}\left(\int_z^{+\infty}b(z)\,|u'(z)|^2\,dz\right)^{1/2}\,.
\eeqa

Since $\|u\|_{\E_0}<+\infty$, for $\varepsilon>0$, there exists $z_0>0$ such that
\beq\label{L}
\int_{z_0}^{+\infty}\Bigl(b(z)\,|u'(z)|^2+a(z)\,|u(z)|^2\Bigr)\,dz\leq\varepsilon.
\eeq

Define $\chi(z)$ on $[0, +\infty)$ by
\beqn \chi(z)=\cases{1\ &\mbox{if}\quad $0\leq z\leq z_0$\\
\ln\left(\frac{\beta_0(z)}{\beta_0(z_1)}\right)\ &\mbox{if}\quad $z_0\leq z\leq z_1$\\
0\ &\mbox{if}\quad $z_1\leq z\leq +\infty$}\nonumber\eeqn
with $z_1$ given by the equation $\displaystyle{\beta_0(z_1)=\rme^{-1}\,\beta_0(z_0)}$. Then we have
\begin{eqnarray*}\fl
\|u-u\,\chi\|_{\E_0}^2&\leq & \int_{z_0}^{+\infty}a(z)\,\Bigl(1-\chi(z)\Bigr)^2|u(z)|^2\,dz +\int_{z_0}^{+\infty}b(z)\,\Bigl(1-\chi(z)\Bigr)^2\,|u'(z)|^2\,dz\nn
 &+&\int_{z_0}^{+\infty}b(z)\,\chi'(z)^2\,|u(z)|^2\,dz\,. \eeqan
The first two terms are small by (\ref{L}), and for the third one, we have from (\ref{acotphi}) and (\ref{L})
 \beqan\int_{z_0}^{+\infty}b(z)\,\chi'(z)^2\,|u(z)|^2\,dz&\leq&\int_{z_0}^{z_1} \frac1{b(z)\ \beta_{0}(z)^2}\,|u(z)|^2\,dz\nn
 &\leq&\varepsilon\int_{z_0}^{z_1} \frac1{b(z)\ \beta_{0}(z)}\,dz\nn
 &\leq& C\varepsilon\,.
\end{eqnarray*}
Since $u\,\chi \in\E_{0,c}$, the density of $\E_{0,c}$ in $\E_0$ is proved.

 For the case when $\dis{\int^{+\infty}\frac1{b(z)}\ dz=+\infty}$, given $z'>0$ we define $\displaystyle{\beta_{0}(z)=\int_{z'}^z\frac1{b(s)}\ ds}$, and we choose $z^*, z$ such that $z'\leq z^*\leq z$. We have
\beqn
|u(z)-u(z^*)|\leq\int_{z^*}^z|u'(s)|\,ds\leq \left(\int_{z^*}^{+\infty}b(s)\,|u'(s)|^2\,ds\right)^{\frac12}\sqrt{\beta_0(z)},
\eeqn
hence
\beqn
|u(z)|\leq |u(z^*)|+ \left(\int_{z^*}^{+\infty}b(s)\,|u'(s)|^2\,ds\right)^{\frac12}\sqrt{\beta_0(z)}.
\eeqn
 This implies
\beq\label{lim}
\lim_{z \to +\infty}\frac{|u(z)|}{\sqrt{\beta_0(z)}}=0\,.
\eeq

Now, by (\ref{lim}) for any $\varepsilon>0$, there exists $z_0>0$ such that
\beqn 
\frac{|u(z_0)|^2}{\beta_0(z_0)}+\int_{z_0}^{+\infty}\Bigl(b(z)\,|u'(z)|^2+a(z)\,|u(z)|^2\Bigr)\,dz\leq\varepsilon.
\eeqn
Then,
\beqn
|u(z)|\leq |u(z_0)|+ \sqrt{\varepsilon}\sqrt{\beta_0(z)},
\eeqn
when $z\geq z_0$. We define $\chi(z)$  by
\beqn \chi(z)=\cases{1\ &\mbox{if}\quad $0\leq z\leq z_0$\\
\ln\left(\frac{\beta_0(z_1)}{\beta_0(z)}\right)\ &\mbox{if}\quad $z_0\leq z\leq z_1$\\
0\ &\mbox{if}\quad $z_1\leq z\leq +\infty$}\eeqn
with $z_1$ given by the equation $\displaystyle{\beta(z_1)=\rme\,\beta_0(z_0)}$,
and we can prove, as above, that there exists a constant $C$ such that
\beqn
\|u-u\chi\|_{\E_0}^2\leq C\,\varepsilon\,.
\eeqn
Thus, in this case also, $\E_{0,c}$ is dense in $\E_0$.\cuad

\begin{lem}\label{pirulo}
\begin{enumerate}
\item The set of all $u\in\E_0$ which vanishes in some neighbourhood of $0$ (depending on $u$) is dense in $\E_0$ if and only if $\displaystyle{\int_0\left(\frac1{b(z)}+a(z)\right)\,dz=+\infty}$

\item $\E_{0,0}$ is dense in $\E$ if and only if $\displaystyle{\int^{\infty}\left(\frac1{b(z)}+a(z)\right)\,dz=+\infty}$ and \\ $\displaystyle{\int_0\left(\frac1{b(z)}+a(z)\right)\,dz=+\infty}$.
\end{enumerate}
\end{lem}

\noindent{\bf Proof:}

{\it (i)} We consider the transformation  $\displaystyle{\phi(z)= \frac1z:(0,+\infty)\to (0,+\infty)}$, and let

\beqn\fl
\E_{\phi}=\Bigl\{u\in H^1_{loc}((0, +\infty): \|u\|_{\phi}^2=\int_0^{+\infty}\Bigl(b_{\phi}(z)\,|u'(z)|^2+a_{\phi}(z)\,|u(z)|^2\Bigr)\, dz <+\infty\Bigr\},
\eeqn
where $\dis {b_{\phi}(z)=z^2\,b\left(1/z\right)}$  and ${\dis a_{\phi}(z)= a\left(1/z\right)/{z^2}}$.

 Then  $\E_{\phi}$ and $\E_0$ are isomorphic,  through the application $\Phi: \E_0\rightarrow \E_{\phi}$ given by $\Phi(v)=u=v\circ \phi$.

 By lemma \ref{densidad1},  $\E_{\phi,c}$ is dense in $\E_{\phi}$ if and only if $\dis \int^{+\infty}\left(\frac1{b_{\phi}(z)}+a_{\phi}(z)\right)\,dz
 =\int_0\left(\frac1{b(z)}+a(z)\right)\,dz=\infty$, and we observe that   $v\in \E_0$ vanishes in a neighbourhood of $0$ if and only if $\Phi(v) \in \E_{\phi,c}$.

 {\it (ii)} follows directly from both assertion {\it (i)} and lemma \ref{densidad1}.\cuad

In this step we have done the assumption that $d=0$ and $k =0$. When $d$ or $k$ are not vanishing, then it suffices to replace $a(z)$ by $a(z)+d(z)+\lambda_kc(z)$  to obtain the appropriate versions of lemmas 3.4 and 3.5.

\subsection*{\bf Step 3: conclusion in the one dimensional case}

\begin{lem}\label{densidadenE0}\
\begin{enumerate}
\item $C_c^{\infty}\left(\rule{0pt}{10.5pt}[\,0,+\infty)\right)\cap \E_0$ is dense in $\E_0$ if and only if $\displaystyle{\int^{+\infty}\left(\frac1{b(z)}+a(z)\right)\,dz=+\infty}$,

\item $C_0^{\infty}\left(\rule{0pt}{10.5pt}(0,+\infty)\right)$ is dense in $\E_0$ if and only if $\displaystyle{\int^{+\infty}\left(\frac1{b(z)}+a(z)\right)\,dz=+\infty}$ and $\displaystyle{\int_0\left(\frac1{b(z)}+a(z)\right)\,dz=+\infty}$.
\end{enumerate}
\end{lem}
{\bf Proof.}

{\it (ii)}
Assume first $C_0^{\infty}\left(\rule{0pt}{10.5pt}(0,+\infty)\right)$ is dense in $\E_0$, then $\E_{0,0}$ must be dense too, and this implies, by lemma \ref{pirulo}, $\displaystyle{\int^{+\infty}\left(\frac1{b(z)}+a(z)\right)\,dz=\int_0\left(\frac1{b(z)}+a(z)\right)\,dz=+\infty}$.

Reciprocally, if $\displaystyle{\int^{+\infty}\left(\frac1{b(z)}+a(z)\right)\,dz=\int_0\left(\frac1{b(z)}+a(z)\right)\,dz=+\infty}$, by lemma \ref{pirulo}, $\E_{0,0}$ is dense in $\E_0$. Therefore it suffices to prove that $C_0^{\infty}\left(\rule{0pt}{10.5pt}(0,+\infty)\right)$ is dense in $\E_{0,0}$. For this purpose we will show that for any  compact interval  $I=[z_0,z_1] \subset (0, +\infty)$, $C_0^{\infty}(I)$ is dense in $\E_I=\{ u\in \E_{0}:\ \mbox{supp}\, {u}\subset I\}.$

Let $\displaystyle{m=\int_I b(z)\,dz}$ and define $\phi: I\to J=[0,m]$ by $\displaystyle{\phi(z)=\int_{z_0}^{z}b(s)\,ds}$. Then, $L^2\left(\rule{0pt}{10.5pt}I,b(z)dz\right)$ and $L^2(J,ds)$ are isomorphic through the application $\Phi: L^2(J,ds)\rightarrow L^2\left(\rule{0pt}{10.5pt}I,b(z)dz\right)$ such that $\Phi(v)=v\circ \phi$.

Let $u\in \E_I$, and denote $f=u'$ and $g=f\circ \phi^{-1}$, $g\in L^2(J,ds)$, then there exists a sequence $(g_n)_{n\geq 0}$ such that $g_n\in C_0(\mathring{J})$\footnote {$C_0(\mathring{J})$ is the space of continuous functions with compact support in $(0,m)$. } for all $n\geq 0$ and $g_n\to g$ in $L^2(J,ds)$. Let $f_n=g_n\circ \phi$, then $f_n \in C_0(\mathring{I})$ and $f_n\to f$ in $L^2\left(\rule{0pt}{10.5pt}I,b(z)dz\right)$, we also have that
\beqn
\int_I\left|\rule{0pt}{10pt}f(z)-f_n(z)\right|\,dz \leq C\left(\int_Ib(z)\,\left|\rule{0pt}{10pt}f(z)-f_n(z)\right|^2dz \right)^{\frac12},
\eeqn
by Cauchy-Schwarz inequality and because $\displaystyle{\frac{1}{b}\in L^1_{loc}\left(\rule{0pt}{10.5pt}(0,\infty)\right)}$. Since $\dis\int_If(z)\,dz=0$ , we deduce that
\beqn
\lim_{n\to\infty}\int_If_n(z)\,dz=0.
\eeqn
Choose $\chi\in C_0(\mathring{I})$, such that $\dis \int_I\chi(z)\,dz=1$, and define
\beqn
\tilde{f}_n=f_n-\left(\int_If_n(z)dz\right)\chi.
\eeqn
Then $\dis \int_I\tilde{f}_n(z)\,dz=0$, $\tilde{f}_n \in C_0(\mathring{I})$ and $\tilde{f}_n\to f$ in $L^2\left(\rule{0pt}{10.5pt}I,b(z)dz\right)$:
\beqa\label{tilde}
\fl\int_I b(z)\,\Bigl(f(z)-\tilde{f}_n(z)\Bigr)^2\,dz&\leq \int_Ib(z)\,\Bigl(f(z)-f_n(z)\Bigr)^2 dz+ \left(\int_If_n(z)\,dz\right)^2\int_I b(z)\,\chi(z)^2\,dz\nn
& \underset{n\to \infty}{\longrightarrow}0\,.
\eeqa
 Set
\beqn
\tilde{u}_n(z)=\int_{z_0}^z\tilde{f}_n(s)\,ds,
\eeqn
since $\dis \int_I\tilde{f}_n(z)\,dz=0$, $\tilde{u}_n(z) \in C_0^1(\mathring{I})$ for all $n\geq 0$, and by (\ref{tilde}),
\beqn
\lim_{n\to\infty}\int_Ib(z)\,\Bigl|u'(z)-\tilde{u}'_n(z)\Bigr|^2dz=0,
\eeqn
and
\beqn
\lim_{n\to\infty}\left\|u-\tilde{u}_n\right\|_{\infty}=0
\eeqn
because
\beqn
\lim_{n\to\infty}\int_I\left|f(z)-\tilde{f}_n(z)\right|\,dz=0.
\eeqn
Hence we have
\beqn
\lim_{n\to\infty}\int_Ia(z)\,\Bigl|u(z)-\tilde{u}_n(z)\Bigr|^2\,dz=0,
\eeqn
so that, finally,
\beqn
\lim_{n\to\infty}\|u-\tilde{u}_n\|_{\E_I}=0.
\eeqn
This proves the density of $C_0^1(I)$ in $\E_I$. To pass from $C_0^1(I)$ to $C_0^{\infty}(I)$, a classical regularization procedure is enough: it shows that $C_0^{\infty}(I)$ is dense in $C_0^1(I)$ for the topology given by the norm
\beqn
\sup_{z\in I}|u(z)|+ \sup_{z\in I}|u'(z)| ;
\eeqn
since $a$ and $b$ are integrable on $I$, this implies the same density for the topology induced by $\E_I$, and part {\it (ii)} of the lemma is completely proved.

Regarding part {\it (i)}, we will be sketchy. The necesity of the condition $\displaystyle{\int^{+\infty}\left(\frac1{b(z)}+a(z)\right)\,dz=+\infty}$ follows from lemma \ref{densidad1}. Its sufficiency needs only to be proved when $C_0^{\infty}\left(\rule{0pt}{10pt}(0,\infty)\right)$ is not dense, that is to say when $\displaystyle{\int_0\left(\frac1{b(z)}+a(z)\right)\,dz < +\infty}$.

But then, the same proof as above works, even when $I=[0, z_1]$.\cuad

\subsection*{\bf Proof of  theorem \ref{q.e.s.a.}}\

Let us now prove theorem \ref{q.e.s.a.} {\it(ii)}:
if $C_0^{\infty}(\O)$ is dense in $\E$,  by lemma \ref{Ek}  $C_0^{\infty}\left(\rule{0pt}{10pt}(0,+\infty)\right)$ is dense in $\E_k$  for all $k \geq 0$, in particular for $k=0$, then by lemma \ref{densidadenE0}, we have $\dis\int^{+\infty}\left(\frac1{b(z)}+d(z)+a(z)\right)dz=\int_0\left(\frac1{b(z)}+d(z)+a(z)\right)dz=+\infty$.

Conversely, if $\displaystyle{\int^{+\infty}\left(\frac1{b(z)}+d(z)+a(z)\right)dz=\int_0\left(\frac1{b(z)}+d(z)+a(z)\right)dz=+\infty}$, we also have $${\int^{+\infty}\left(\frac1{b(z)}+d(z)+a(z)+\lambda_k\,c(z)\right)dz=
\int_0\left(\frac1{b(z)}+d(z)+a(z)+\lambda_k\,c(z)\right)dz=+\infty}\,,$$ then $C_0^{\infty}\left(\rule{0pt}{10pt}(0,+\infty)\right)$ is dense in $\E_k$  for all $k$, we can see it changing $a(z)$ by $d(z)+a(z)+\lambda_k\,c(z)$ in all the previous results, and again by lemma \ref{Ek},  $C_0^{\infty}(\O)$ is dense in $\E$.

The proof of {\it(i)} analogously follows. Theorem \ref{q.e.s.a.} is completely proved.\cuad

\begin{rem}\rm\label{correccion}Under different hypotheses, when the coefficients of the operator $A$ depend on $(z,\mathbf{x})$ we have given a characterization of {\it q.e.s.a.} operators in \cite{GST}.
Warning: in page 21  of that reference,  the integrand of (43)  was mistakenly written as $\frac 1{M_{n+1,n+1}(z,\mathbf{ x})}$ instead of $\dis (M^{-1})_{n+1,n+1}(z, \mathbf{x})$.
\end{rem}

 \subsection*{\bf Essentially selfadjointness  characterization}

The characterization of {\it e.s.a.} for the operator $A$ defined in \eq{operador} will rely on the real-valued solutions of the O.D.E.
\beq\lab{ODE} - \Bigl(b(z)\, u'(z)\Bigr)'+d(z)\, u(z)=0\eeq
on $(0,z')$ and on $(z',+\infty)$.

A typical case is when $\dis \int_0 a(z)\, dz<+\infty$, but $\dis \int^{+\infty} a(z)\, dz=+\infty$. Then since we may assume $A$ to be {\it q.e.s.a.} (otherwise it cannot be {\it e.s.a.}), we have  $\dis \int_0\left( \frac 1{b(z)}+d(z)\right)\, dz=+\infty$. In such a case, we will show that there is a unique solution of \eq{ODE}, denoted by $\alpha$, such that
\beqa\lab{alfa1}\cases{ \alpha\ \mbox{is a solution of \eq{ODE} in } (0,z'),\\\alpha(z')=1,\\ \int_0^{z'} \Bigl(b(z)\,\alpha'(z)^2+d(z)\, \alpha(z)^2\Bigr)\, dz<+\infty\ .}\eeqa
Then, we define $\beta(z)$, $z \in(0,z')$, by
\beq\lab{beta1} \beta(z)=\alpha(z)\,\int_z^{z'} \frac1{b(s)\,\alpha(s)^2}\ ds\,.\eeq
Note that, by construction, $\beta$ is another solution of \eq{ODE} in $ (0,z')$. We shall prove that: $A$ is {\it e.s.a.} if and only if $\dis \int_0 \beta(z)^2\,a(z)\, dz=+\infty$.

In the case where the role of $0$ and $+\infty$ are exchanged, the result is similar. We will show that there exists a unique function $\alpha$ such that
\beqa\lab{alfa2}\cases{ \alpha(z)\ \mbox{is a solution of \eq{ODE} in } (z',+\infty),\\\alpha(z')=1,\\ \int_{z'}^{+\infty} \Bigl(b(z)\,\alpha'(z)^2+d(z)\, \alpha(z)^2\Bigr)\, dz<+\infty\ .}\eeqa
Then, we define $\beta(z)$, $z \in(z',+\infty)$, by
\beq\lab{beta2} \beta(z)=\alpha(z)\,\int_{z'}^z \frac1{b(s)\,\alpha(s)^2}\ ds\ ,\eeq
and we shall prove that: $A$ is {\it e.s.a.} if and only if $\dis \int^{+\infty} \beta(z)^2\,a(z)\, dz=+\infty$.

Note that, when $d(z)\equiv0$ the problem considerably simplifies  since, in this case, $\alpha\equiv1$ and $\beta(z)$ turns out to be either $\dis \beta_0(z)=\int_z^{z'} \frac 1{b(z)}\, dz$ or $\dis \beta_0(z)=\int_{z'}^z \frac 1{b(z)} dz$ respectively.

\begin{notation}
\rm We denote $\left(\rule{0pt}{10pt}\alpha(z),\beta(z)\right)$ the above couples of solutions of \eq{ODE}; the context will indicate whether  $z \in(0,z')$, in which case $\left(\rule{0pt}{10pt}\alpha(z),\beta(z)\right)$ are given by \eq{alfa1} and \eq{beta1}, or $z \in(z',+\infty)$, where $\left(\rule{0pt}{10pt}\alpha(z),\beta(z)\right)$ are given by \eq{alfa2} and \eq{beta2}.
\end{notation}

With this notation, the result is the following.

\begin{theorem}\lab{esa} Assume the operator $A$ given in \eq{operador} to be {\it q.e.s.a.}, that is to say
\beqn\fl \int_0\left( \frac 1{b(z)}+d(z)+a(z)\right)\, dz=\int^{+\infty}\left( \frac 1{b(z)}+d(z)+a(z)\right)\, dz=+\infty \ . \eeqn
There are four cases:

\begin{enumerate}

\item If $\dis \int_0 a(z) \,dz= \int^{+\infty} a(z) \,dz=+\infty$, then  $A$ is {\it e.s.a.};\vs {.3}

\item If $\dis  \int_0 a(z) \,dz<+\infty$ and $\dis  \int^{+\infty} a(z) \,dz=+\infty$, then  $A$ is {\it e.s.a.} if and only if $\dis \int_0 \beta(z)^2 a(z) \,dz= +\infty$ ;\vs {.3}

\item If $\dis  \int_0 a(z) \,dz=+\infty$ and $\dis  \int^{+\infty} a(z) \,dz<+\infty$, then  $A$ is {\it e.s.a.} if and only if $\dis \int^{+\infty} \beta(z)^2 a(z) \,dz= +\infty$ ;\vs {.3}

\item If $\dis  \int_0 a(z) \,dz<+\infty$ and $\dis  \int^{+\infty} a(z) \,dz<+\infty$, then  $A$ is {\it e.s.a.} if and only if $\dis \int_0 \beta(z)^2 a(z) \,dz=\int^{+\infty} \beta(z)^2 a(z) \,dz= +\infty$ .
\end{enumerate}

\end{theorem}

\begin{rem} \rm
 Take care of the uniqueness of $\alpha$ (and thus the meaningfulness of the definitions above): it holds when  $\dis \int_0\Bigl( \frac 1{b(z)}+d(z)\Bigr)\, dz=+\infty$ or  $\dis \int^{+\infty} \Bigl( \frac 1{b(z)}+d(z)\Bigr)\, dz=+\infty$, according to where the variable $z$ lives.
\end{rem}

\subsection*{\bf Preliminary step: study of solutions of \eq{ODE}}

\begin{lem}\lab{lI} Let $u(z)$ be a solution of \eq{ODE} in some interval $I\subset (0, +\infty)$. Then the function $b(z)\,u(z)'\,u(z)$ is increasing in $I$.
\end{lem}
\noindent{\bf Proof:}

From \eq{ODE} we obtain $$-\Bigl(b(z)\,u(z)'\,u(z)\Bigr)'+b(z)\,u'(z)^2+d(z)\,u(z)^2=0\,,$$ showing that  $\Bigl(b(z)\,u(z)'\,u(z)\Bigr)'$ is nonnegative.\cuad

\begin{lem} \lab{lII}Let $u(z)$ be a solution of \eq{ODE} in  $(0,z')$. Then
\beqn \int_0^{z'} \Bigl(b(z)\,u'(z)^2+d(z)\,u(z)^2\Bigr)\,dz=+\infty\eeqn
if and only if
\beqn \lim_{z\to 0^+} b(z)\,u'(z)\,u(z)= -\infty\,.\eeqn
\end{lem}
\noindent{\bf Proof:}

Since $u(z')$ and $u'(z')$ exist, the proof follows immediately from the fact that, for $0<z_0<z'$,  we have
\beqan\fl \int_{z_0}^{z'}  \Bigl(b(z)\,u'(z)^2+d(z)\,u(z)^2\Bigr)\,dz&=\int_{z_0}^{z'}  \Bigl(b(z)\,u(z)'\,u(z)\Bigr)'\,dz\nn &=b(z')\,u'(z')\,u(z')-b(z_0)\,u'(z_0)\,u(z_0)\,.\eeqan\cuad

\begin{lem}\lab{lIII}\ Let $z'>0$ be chosen.
\begin{enumerate}
\item There exists at least one solution $\alpha(z)$ of \eq{ODE}, in the interval $(0,z')$, such that
\beqn \alpha(z')=1 \eeqn
and
\beqn \int_0^{z'}\Bigl(b(z)\,\alpha'(z)^2+d(z)\,\alpha(z)^2\Bigr)\,dz<+\infty\,.\eeqn

This solution is positive and increasing in $(0,z')$, satisfying
\beqn \lim_{z\to 0^+} b(z)\,\alpha'(z)\,\alpha(z)= 0\,.\eeqn
\item If in addition {$\dis \int_0^{z'}\left( \frac 1{b(z)}+d(z)\right)\, dz=+\infty$}, this solution is unique.
\end{enumerate}

\end{lem}
\noindent{\bf Proof:}

Let  $L_b^2\left(\rule{0pt}{10.5pt}(0,z')\right)$ be the space of measurable functions $f(z)$ such that
\beqn  \int_0^{z'} b(z)\, f(z)^2\, dz<+\infty\,. \eeqn
We define, for any $f$ in this space, the function $Tf$ by
\beqn Tf(z)=1-\int_z^{z'} f(s)\, ds\,,\eeqn
so that $Tf\in C((0,z')\bigr)\bigcap H_{loc}^1\left(\rule{0pt}{10.5pt}(0,z')\right)$, with $(Tf)'(z)=f(z)$. Let
\beqn q(f)=\int_0^{z'}\left(b(z)\,f(z)^2+d(z)\,\Bigl(Tf(z)\Bigr)^2\right)\,dz ,\eeqn
 taking values in $(0,+\infty]$, and
\beqn q_0=\inf_{f\in L_b^2}\,q(f)\,.\eeqn
Note that $q_0$ is finite since, for example, for $\dis f(z)= \frac 1{z'-z_0} \mathbf 1_{[z_0,z']}(z)$ for some $0<z_0<z'$, $q(f)<+\infty$.
We shall show that $q_0$ is in fact a minimum. To this end, let $(f_n)_{n\in\N}$ be a minimising sequence
\beqn \lim_{n\to+\infty}\,q(f_n)=q_0\,.\eeqn
Then, by construction,
\beqn \sup_{n\in\N}\,\left\|f_n\right\|_{L_b^2}<+\infty\,,\eeqn
so that (up to extracting a subsequence) we may suppose that the sequence $(f_n)$ has a weak limit $f_0$ in $L_b^2\left(\rule{0pt}{10.5pt}(0,z')\right)$. Let us prove that $q(f_0)=q_0$.

For any $z_0\in(0,z')$ and for all
$z\geq z_0$
\beqn\left|Tf_n(z)\right|\leq 1+\left(\int_{z_0}^{z'} \frac 1{b(z)}\  dz\right)^{1/2}\,\left\|f_n\right\|_{L_b^2}\leq C(z_0)\,, \eeqn
and \beqn Tf_0(z)\,\mathbf 1_{[z_0,z']}(z)=\lim_{n\to+\infty}\,Tf_n(z)\,\mathbf 1_{[z_0,z']}(z)\,.\eeqn
So, by dominated convergence, we have
\beqn \lim_{n\to+\infty}\int_{z_0}^{z'} d(z)\,\Bigl(Tf_n(z)\Bigr)^2\,dz=\int_{z_0}^{z'} d(z)\,\Bigl(Tf_0(z)\Bigr)^2\,dz\,.\eeqn
Also we know that
\beqn \int_{z_0}^{z'} b(z)\,f_0(z)^2\,dz\leq\liminf_{n\to+\infty}\int_{z_0}^{z'} b(z)\,f_n(z)^2\,dz\,,\eeqn
since $f_0=  \underset{n\to+\infty}{\mbox{w-lim}} \, f_n$ in  $L_b^2\left(\rule{0pt}{10.5pt}(0,z')\right)$ as well. From these two facts, we deduce
\beqan \fl\int_{z_0}^{z'} \left(b(z)\,f_0(z)^2+d(z)\,\Bigl(Tf_0(z)\Bigr)^2\right)\,dz\nn\leq\liminf_{n\to+\infty}\int_{z_0}^{z'} \left(b(z)\,f_n(z)^2+d(z)\,\Bigl(Tf_n(z)\Bigr)^2\right)\,dz\leq q_0\,.\eeqan
Letting $z_0\to 0^+$, we obtain $q(f_0)\leq q_0$, and thus $q(f_0)= q_0$ as desired.

Let now $\alpha(z)=Tf_0(z)$. For any $u\in C\left(\rule{0pt}{10.5pt}(0,z')\right)\bigcap H_{loc}^1\left(\rule{0pt}{10.5pt}(0,z')\right)$, with $u(z')=1$, define
\beqan Q(u)&=q(u')\nn&=\int_0^{z'} \Bigl(b(z)\,u'(z)^2+d(z)\,u(z)^2\Bigr)\,dz \,.  \eeqan
We have proved that
\beqn Q(\alpha)=\min_u\,Q(u)\,.\eeqn
We define
$$
\alpha^+ =\left\{\begin{array}{cc}
 \alpha &   \mbox{if}\ \alpha \geq 0\\
 0  &  \mbox{if not}
\end{array}\right. \hs 1
\text{and} \hs 1
\alpha^- =\left\{\begin{array}{cc}
 -\alpha &   \mbox{if}\ \alpha \leq 0\\
 0  &  \mbox{if not}
\end{array}\right.,$$
so that
$\alpha=\alpha^+-\alpha^-$
and
$\alpha^+ \alpha^-  =  0.$
Then we have that  $Q(\alpha^+)\leq Q(\alpha)$ with strict inequality if and only if $\alpha^-\neq0$, and since $\alpha^+\in C\left(\rule{0pt}{10.5pt}(0,z')\right)\bigcap H_{loc}^1\left(\rule{0pt}{10.5pt}(0,z')\right)$ with $\alpha^+(z')=1$, we must have
\beqn Q(\alpha^+)=Q(\alpha)\,,\eeqn
and $\alpha^+=\alpha$, i.e., $\alpha$ is positive in $(0,z']$.

If $\psi\in C\left(\rule{0pt}{10.5pt}(0,z')\right)\bigcap H_{loc}^1\left(\rule{0pt}{10.5pt}(0,z')\right)$ is such that $Q(\alpha+t\psi)<+\infty$ for all $t\in \R$ and $\psi(z')=0$, we must have
\beqn Q(\alpha)\leq Q(\alpha+t\psi)\,,\eeqn
and this implies
\beqn \int_0^{z'} \Bigl(b(z)\,\alpha(z)'\psi(z)'+d(z)\,\alpha(z)\,\psi(z)\Bigr)\,dz=0 \,.  \eeqn
This, in particular, is true for all $\psi\in C_0^{\infty}\left(\rule{0pt}{10.5pt}(0,z')\right)$, implying that
\beqn - \Bigl(b(z)\, \alpha(z)'\Bigr)'+d(z)\, \alpha(z)=0\eeqn
in $(0,z')$.

But then, this means that
\beqn \int_0^{z'} \Biggl(b(z)\,\alpha(z)'\psi(z)'+\Bigl(b(z)\,\alpha(z)'\Bigr)'\,\psi(z)\Biggr)\,dz=0   \eeqn
for all $\psi\in C\left(\rule{0pt}{10.5pt}(0,z')\right)\cap H_{loc}^1\left(\rule{0pt}{10.5pt}(0,z')\right)$ with $Q(\alpha+t\psi)<+\infty$ and $\psi(z')=0$. Therefore
\beqn \lim_{z\to 0^+} b(z)\,\alpha'(z)\,\psi(z)= 0\,.\eeqn
Choosing $\psi=\alpha\,\eta$, where $\eta\in C^\infty\left(\rule{0pt}{10.5pt}0,+\infty)\right)$, $\eta=1$ near $0$ and $\eta=0$ near $z'$, we get
\beqn \lim_{z\to 0^+} b(z)\,\alpha'(z)\,\alpha(z)= 0\,.\eeqn
With lemma \ref{lI}, this shows that (recall that $\alpha$ is positive) $\alpha^2$ and hence $\alpha$ are both increasing in $(0,1)$. Thus, part {\it (i)} is entirely proved.

{\it (ii)} Let
\beqn \beta(z)=\alpha(z)\,\int_z^{z'} \frac1{b(s)\,\alpha(s)^2}\ ds\ .\eeqn
Then, $\beta(z)$ is another solution of \eq{ODE} in $(0,z')$, so that any solution writes $\lambda\, \alpha(z)+\mu\, \beta(z)$, $\lambda,\mu\in\R$. The uniqueness of $\alpha(z)$ will follow from the proof of
\beq \lab{end16} \int_0^{z'} \Bigl(b(z)\,\beta'(z)^2+d(z)\,\beta(z)^2\Bigr)\,dz=+\infty\,.\eeq
A direct calculation shows that $\beta(z')=0$ and $\beta'(z')=\dis-\frac 1{ b(z')}$. Thus, from the O.D.E. \eq{ODE}, we obtain
\beqn -\beta'(z)=\frac 1{b(z)}+\frac 1{b(z)}\,\int_z^{z'} d(s)\,\beta(s)\ ds\ .\eeqn
Since $\beta$ is positive by construction, it turns out to be decreasing in $(0,z')$, with
\beqn \left|\beta'(z)\right|\geq\frac 1{b(z)}\ , \hs 1 0<z\leq z'\,,\eeqn
and
\beq \lab{p17}\beta(z)\geq\int_z^{z'} \frac 1{b(s)}\,ds\,=:\beta_0(z)\,.\eeq
Hence, there exists a constant $C$ such that $\beta(z)\geq C$ if $\dis z\leq z'/2$, and we obtain
\beqan \int_0^{z'} \Bigl(b(z)\,\beta'(z)^2+d(z)\,\beta(z)^2\Bigr)\,dz &\geq\int_0^{z'} \frac 1{b(z)}\, dz+C^2\int_0^{z'/2}d(z)\, dz\nn&=+\infty\,.\eeqan
The lemma is proved.\cuad

Lemma \ref{lIII} has an analogous counterpart near $+\infty$, which is the following.

\begin{lem}\lab{lIV}\ Let $z'>0$ be chosen.
\begin{enumerate}
\item There exists at least one solution $\alpha(z)$ of \eq{ODE}, in the interval $(z',+\infty)$, such that
\beqn \alpha(z')=1 \eeqn
and
\beqn \int_{z'}^{+\infty} \Bigl(b(z)\,\alpha'(z)^2+d(z)\,\alpha(z)^2\Bigr)\,dz<+\infty\,.\eeqn

This solution is positive and decreasing in $(z',+\infty)$, satisfying
\beqn \lim_{z\to +\infty} b(z)\,\alpha'(z)\,\alpha(z)= 0\,.\eeqn
\item If in addition {$\dis \int^{+\infty}\left( \frac 1{b(z)}+d(z)\right)\, dz=+\infty$}, this solution is unique.
\end{enumerate}
\end{lem}

\noindent{\bf Proof:}

 By making the change of variable $\dis z\mapsto \frac {z'^2}z$, the proof immediately follows from the previous lemma.\cuad

\begin{rem}\rm The function $\alpha(z)$ given in $(0,z')$ (respectively in $(z',+\infty)$) by lemma \ref{lIII} (resp. lemma \ref{lIV}) is not a solution of \eq{ODE} on $(0,+\infty)$, but of
\beqn - \Bigl(b(z)\, \alpha'(z)\Bigr)'+d(z)\, \alpha(z)=\lambda\, \delta_{z'}(z) \,,\eeqn
where $\delta_{z'}(z)$ is the Dirac measure at $z=z'$, and $\dis \lambda=\int_0^{+\infty} \Bigl(b(z)\,\alpha'(z)^2+d(z)\,\alpha(z)^2\Bigr)\,dz$.
\end{rem}

\subsection*{\bf Main step: {\it e.s.a.} characterization  in dimension one}

Let us consider now the operator
\beqn A_0\,u(z)=\frac{1}{a(z)}\,\Biggl(-\Bigl(b(z)\, u'(z)\Bigr)'+d(z)u(z)\Biggr)\eeqn
defined as in \eq {Ak} with
\beq\lab{top18} \int_{0}\left( \frac 1{b(z)}+d(z)\right)\, dz=\int^{+\infty}\left( \frac 1{b(z)}+d(z)\right)\, dz=+\infty
\,. \eeq

\begin{lem}\lab{lV}
If $\dis  \int_{0} a(z)\, dz=\int^{+\infty}a(z)\, dz=+\infty$, $A_0$ is an  {\it e.s.a.} operator.
\end{lem}

\noindent{\bf Proof:}

  Assume $A_0$ is not {\it e.s.a.}. By lemma \ref{equivalente} there exists $u\in H_0$ such that
\beqn -\Bigl(b(z)\,u'(z)\Bigr)'+d(z)\,u(z)+a(z)\,u(z)=0\,,\eeqn
and $u\notin \E_0$, i.e., either $\dis  \int_0 \Bigl(b(z)(u'(z))^2+\Bigl(d(z)+a(z)\Bigr)\,u(z)^2\Bigr)\,dz=+\infty$ or $\dis  \int^{+\infty} \Bigl(b(z)(u'(z))^2+\Bigl(d(z)+a(z)\Bigr)\,u(z)^2\Bigr)\,dz=+\infty$ (or both).

If $\dis  \int_0 \Biggl(b(z)\,u'(z)^2+\Bigl(d(z)+a(z)\Bigr)\,u(z)^2\Biggr)\,dz=+\infty$, by lemma \ref{lII} (changing $d$ in $d+a$) we have
\beqn \lim_{z\to 0^+} b(z)\,u'(z)\,u(z)= -\infty\,.\eeqn
In particular, $u'(z)\,u(z)<0$ for $z\leq z_0$, for some $z_0>0$, so that $u^2$ is decreasing in $(0,z_0]$. But since $\dis  \int_0^{+\infty} a(z)\, u(z)^2\,dz<+\infty$, this implies  $\dis  \int_{0} a(z)\, dz<+\infty$, which is a contradiction.

If $\dis  \int^{+\infty} \Biggl(b(z)\,u'(z)^2+\Bigl(d(z)+a(z)\Bigr)\,u(z)^2\Biggr)\,dz=+\infty$, a change of variable reduces the proof to the preceding case.\cuad

\begin{lem}\lab{lVI}
Assume $\dis  \int_{0} a(z)\, dz<+\infty$ and $\dis  \int^{+\infty}a(z)\, dz=+\infty$. Then, $A_0$ is an {\it e.s.a.} operator if and only if $\dis  \int_{0}\beta(z)^2 a(z)\, dz=+\infty$.
\end{lem}

\noindent{\bf Proof:}

We first assume that $\dis  \int_{0}\beta(z)^2 a(z)\, dz<+\infty$. We set $u(z)=\beta(z)\,\eta(z)$  with  $\eta\in C^\infty\left(\rule{0pt}{10pt}[0,+\infty)\right)$, $\eta=1$ near $0$ and $\eta=0$ for $z\geq\varepsilon$. Then $u\in H_0$ and $A_0^*u\in H_0$. But by the hypotheses \eq{top18},  $u\notin \E_0$  (see \eq{end16} in the proof of lemma \ref{lIII}). Thus $A_0$ is not {\it e.s.a.}.

Reciprocally, assume that $A_0$ is not {\it e.s.a.}. Then there exists $u\in  H_0$ such that
\beqn -\Bigl(b(z)\,u'(z)\Bigr)'+d(z)\,u(z)+a(z)\,u(z)=0\,,\eeqn
and $u\notin \E_0$.

Since $\dis  \int^{+\infty}a(z)\, dz=+\infty$ and $\dis  \int^{+\infty} u(z)^2 a(z)\, dz<+\infty$, the same argument as in lemma \ref{lV} shows that necessarily
\beqn \int^{+\infty} \Bigl(b(z)\,u'(z)^2+d(z)\,u(z)^2\Bigr)\,dz<+\infty\,.\eeqn
Thus we must have
\beqn \int_0\Bigl(b(z)\,u'(z)^2+d(z)\,u(z)^2\Bigr)\,dz=+\infty\,.\eeqn
By lemma \ref{lII}, $\dis\lim_{z\to 0^+} b(z)\,u'(z)\,u(z)= -\infty$, and in particular, $u^2$  is decreasing in $(0,z_0)$ for some $z_0>0$. We may assume that $u(z_0)>0$ and $u'(z_0)<0$ (up to changing $u$ in $-u$). Let $C_1$ and $C_2$ be two constants such that
\beqn\cases{C_1\, \alpha(z_0)+C_2\, \beta(z_0)=u(z_0),\\ C_1\, \alpha'(z_0)+C_2\, \beta'(z_0)=u'(z_0)\, .}\eeqn
They exist because we know that the Wronskian $b(z)\Bigl(\alpha(z)\beta'(z)-\alpha'(z)\beta(z)\Bigr)$ is never vanishing \footnote{In fact, it is a constant, equal to $-1$.}. Moreover, we must have $C_2\neq0$, otherwise $u(z_0)$ and $u'(z_0)$ would have the same sign (recall that $\alpha$  is positive and increasing, by lemma \ref{lIII}). We even have $C_2>0$\footnote{$C_2=b(z_0)[u(z_0)\alpha'(z_0)-u'(z_0)\alpha(z_0)]>0$}.

Let $v(z)=u(z)-C_1\, \alpha(z)-C_2\, \beta(z)$. We have
\beqn -\Bigl(b(z)\,v'(z)\Bigr)'+d(z)\,v(z)+a(z)\,u(z)=0\,,\eeqn
with $v(z_0)=v'(z_0)=0$, $u>0$ in $(0,z_0]$. By classical arguments, $v$ must be positive and decreasing in $(0,z_0]$:

\begin{itemize}
\item It is so in some neighborhood of $z_0$, because $\Bigl(b(z)\, v'(z)\Bigr)'>0$ near $z_0$ and $v'(z_0)=0$, so that $v'(z)<0$ in $(z_0-\epsilon,z_0)$;

\item it cannot change its sense of variation in $(0,z_0)$ ($v(z_1)>0$, $v'(z_1)=0$, $v''(z_1)\leq0$ at some $z_1<z_0$ is impossible).

\end{itemize}

Hence, since  $C_2>0$, we have
\beqn  \beta(z)\leq \frac 1{C_2}\left(u(z)-C_1\,\alpha(z)\right) \eeqn
in $(0,z_0]$. Since $\dis \alpha$  is bounded, $\dis  \int_{0} a(z)\, dz<+\infty$ and $u\in  H_0$, this implies
\beqn \int_{0}\beta(z)^2 a(z)\, dz<+\infty\,,\eeqn
and the proof is finished.\cuad

\begin{lem}\lab{lVII}
Assume $\dis  \int_{0} a(z)\, dz=+\infty$ and $\dis  \int^{+\infty}a(z)\, dz<+\infty$. Then, $A_0$ is an {\it e.s.a.} operator if and only if $\dis  \int^{+\infty}\beta(z)^2\, a(z)\, dz=+\infty$.
\end{lem}

\noindent{\bf Proof:}

The result follows by a change of variable and the preceding lemma.\cuad

\begin{lem}\lab{lVIII}
Assume $\dis  \int_{0} a(z)\, dz<+\infty$ and $\dis  \int^{+\infty}a(z)\, dz<+\infty$. Then, $A_0$ is an {\it e.s.a.} operator if and only if $\dis  \int^{+\infty}\beta(z)^2 a(z)\, dz=+\infty$.
\end{lem}

\noindent{\bf Proof:}

If $A_0$ is not {\it e.s.a.}, there exists  $u\in H_0$ solution of
\beqn -\Bigl(b(z)\,u'(z)\Bigr)'+d(z)\,u(z)+a(z)\,u(z)=0\,,\eeqn
and  either $\dis \int_0 \Bigl(b(z)\,u'(z)^2+d(z)\,u(z)^2\Bigr)\,dz=+\infty$ or $\dis \int^{+\infty} \Bigl(b(z)\,u'(z)^2+d(z)\,u(z)^2\Bigr)\,dz=+\infty$. Use the arguments of lemma \ref{lVI} or lemma \ref{lVII}, depending on the case.

Reciprocally, as we have done in lemma \ref{lVI}, we consider $u(z)=\beta(z)\,\eta(z)$  for a suitable $\eta$ and the result follows.\cuad

\subsection*{\bf Final step: reduction to the one-dimensional case}

Defining the operators $A_k$ as in (\ref{Ak}), i.e.,
 \beqn
A_k\, u(z)=\frac1{a(z)}\Biggl( -\biggl(b(z)u'(z)\Bigr)'+\Bigl(\lambda_k\,c(z)+{d(z)}\Bigr)u(z)\Biggr)\,,
 \eeqn
we have the following result:

 \begin{lem}\lab{X}
$ A$ is an {\it e.s.a.} operator if and only if for all $k\geq 0\ A_k$ is an {\it e.s.a.} operator.
 \end{lem}

\noindent{\bf Proof:}

We use the notation introduced in step 1 of the proof of theorem \ref{q.e.s.a.}.
By Lemma \ref{equivalente}, if $A_k$ is not {\it e.s.a.}, there exists $u\in H_0$, $u\in D(A^*_k)$ but $u\notin \E_k$. This implies that $\varphi=u \otimes \psi_k \in D(A^*)$ and $\varphi \notin \E$, so that $A$ is not {\it e.s.a.} .

Reciprocally, if $A$ is not {\it e.s.a.}, there exists $\varphi \in H$ non vanishing, such that
\beqn A^*\varphi +\varphi=0.\eeqn
Decompose
\beqn\varphi= \sum_{k\geq 0}u_k\otimes \psi_k,\eeqn
there exists $k$ such that $u_k\neq 0$.  If $\phi\in C_0^{\infty}\left(\rule{0pt}{10pt}(0,\infty)\right)$, we have
\begin{eqnarray*}0=<\varphi, A(\phi\otimes\psi_k) + \phi\otimes\psi_k>_H=<u_k,A_k\phi +\phi>_{H_0},
\end{eqnarray*}
which means that $A^*_ku_k+u_k=0$. Thus $A_k$ is not {\it e.s.a.} by lemma \ref{equivalente} again.\cuad

\subsection*{\bf Proof of theorem \ref{esa}}\

{\it (i)} If $\dis \int_0 a(z) \,dz=+\infty$ and $\dis \int^{+\infty} a(z) \,dz=+\infty$, then $\dis \int_0 \Bigl(a(z)+\lambda_k\,c(z)\Bigr) \,dz=+\infty$ and $\dis \int^{+\infty} \Bigl(a(z)+\lambda_k\,c(z)\Bigr) \,dz=+\infty$, for all $k\geq0$. Therefore $A_k$ is {\it e.s.a.} by lemma \ref{lV} with $a$ changed in $a+\lambda_k\,c(z)$, and by lemma \ref{X} $A$  is {\it e.s.a.}.

In the cases {\it (ii)}, {\it (iii)} and {\it (iv)} if $A$ is {\it e.s.a.} it follows by lemma \ref{X} that in particular $A_0$ is {\it e.s.a.}. Then  lemmas \ref{lVI}, \ref{lVII} and  \ref{lVIII} give the result.

For the converse, let us take the case {\it (ii)}. If $A$ is not {\it e.s.a.}, by lemma \ref{X} there exists $k\geq0$ such that $A_k$ is not {\it e.s.a.}. Then by lemma \ref{lVI}

\beq\label{kk} \int_{0}\beta_k(z)^2 a(z)\, dz<+\infty,\eeq
where $\beta_k$ is the solution of
\beqn - \Bigl(b(z)\, u'(z)\Bigr)'+(c(z)\lambda_k+d(z))\, u(z)=0\eeqn
on $(0,z')$ with Cauchy data $u(z')=0$ and $\dis u'(z')=-\frac1{b(z')}$. A classical comparison principle, applied to the functions $\beta_k$ and $\beta$, defined in (\ref{beta1}), give us
$0\leq\beta\leq\beta_k$ on $(0,z').$ Then (\ref{kk}) implies

$$ \int_{0}\beta(z)^2 a(z)\, dz<+\infty \, ,$$
as desired.

The other cases are analogous.

Theorem \ref{esa} is completely proved.\cuad

\begin{rem}\rm
The precise definition of the function $\beta(z)$ is needed only for the sufficiency of the condition
\beqn \int_{0}\beta(z)^2 a(z)\, dz<+\infty\eeqn
for $A$ to be {\it e.s.a.}. This is not used in the reciprocal, where the ``massless-$\beta$\,''
\beqn \beta_0(z)=\int_z^{z'} \frac 1{b(s)}\,ds\,\,\eeqn
would have worked as well (see \eq{p17}). But, for the sufficiency, if we choose $u(z)=\beta_0(z)\,\eta(z)$  in lemma \ref{lVI}, with  $\eta\in C^\infty([0,+\infty))$, $\eta=1$ near $0$ and $\eta=0$ for $\dis z\geq\frac{z'}2$, then
\beqn A_0^*\,u(z)=\frac1{a(z)}\Biggl( -\Bigl(b(z)\,\beta_0(z)\,\eta'(z)\Bigr)' + d(z)\,\beta_0(z)\,\eta(z)\Biggr)\,, \eeqn
and this belongs to $H_0$ only when
\beqn \int_{0}d(z)^2\,\beta_0(z)^2 \frac1{a(z)}\, dz<+\infty\,.\eeqn
\end{rem}

This gives a necessary and sufficient condition for {\it e.s.a.} in terms of $\beta_0(z)$ only, not $\beta(z)$, when $\dis \frac{d(z)}{a(z)}$ is bounded:

\begin{cor}\lab{coro}
When $\dis\frac{ d(z)}{a(z)}$ is bounded near $0$,  $\dis  \int_{0} a(z)\, dz<+\infty$ and $\dis  \int^{+\infty}a(z)\, dz=+\infty$, $A$ is {\it e.s.a.} if and only if $\dis  \int_0\beta_0(z)^2 a(z)\, dz=+\infty$.
\end{cor}

There are similar statements in the other cases.

\bigskip

\begin{rem}\label{01}\rm{\bf The previous results in the domain $(z_0, z_1)\times \M$}

In some relevant examples one is lead to consider $\Omega=(z_0, z_1)\times \M$, $0\leq z_0\leq z_1\leq\infty$, and a differential operator $A$ defined as in (\ref{operador}) by

 \beqn\fl
A\varphi(z, \mathbf{x})=\frac{1}{a(z)}\left\{- \pd z \Bigl(b(z) \pd z \varphi(z, \mathbf{x})\Bigr)- c(z) \Delta_{\mch \M} \varphi(z, \mathbf{x}) + d(z)\varphi(z, \mathbf{x})\right\},
\eeqn
 for all $\varphi \in C_0^{\infty}(\Omega)$,
   where the functions $a$, $b$, and $c$ satisfy the following hypotheses:
 \begin{itemize}
 \item $a$, $c$, $d \in  L_{loc}^1\left(\rule{0pt}{10.5pt}(z_0, z_1)\right)$, $b \in  C\left(\rule{0pt}{10.5pt}(z_0, z_1)\right)$

 \item $a>0$, $b>0$, $c>0$ and $d\geq 0$ in $(z_0,z_1)$
 \item $a^{-1}$, $b^{-1}$, $c^{-1} \in L_{loc}^1\left(\rule{0pt}{10.5pt}(z_0,z_1)\right).$
\end{itemize}

The previous results straightforwardly generalize to such a case. For the convenience of the reader, we state the two main theorems.
\begin{theorem}\lab{qesa1}

$A$ is a q.e.s.a. operator in $H$ if and only if $\dis \int^{z_1}\left(\frac1{b(z)}+d(z)+a(z)\right)dz$ $=+\infty$ and $\displaystyle{\int_{z_0}\left(\frac1{b(z)}+d(z)+a(z)\right)dz=+\infty}$.
\end{theorem}

\begin{theorem}\lab{coresa} We assume $A$ is a {\it q.e.s.a.} operator,
There are four cases:

\begin{enumerate}

\item If $\dis \int_{z_0}a(z) \,dz= \int^{z_1} a(z) \,dz=+\infty$, then  $A$ is {\it e.s.a.};\vs {.3}

\item If $\dis  \int_{z_0} a(z) \,dz<+\infty$ and $\dis  \int^{z_1} a(z) \,dz=+\infty$, then  $A$ is {\it e.s.a.} if and only if $\dis \int_{z_0} \beta(z)^2 a(z) \,dz= +\infty$ ;\vs {.3}

\item If $\dis  \int_{z_0} a(z) \,dz=+\infty$ and $\dis  \int^{z_1} a(z) \,dz<+\infty$, then  $A$ is {\it e.s.a.} if and only if $\dis \int^{z_1} \beta(z)^2 a(z) \,dz= +\infty$ ;\vs {.3}

\item If $\dis  \int_{z_0}a(z) \,dz<+\infty$ and $\dis  \int^{z_1} a(z) \,dz<+\infty$, then  $A$ is {\it e.s.a.} if and only if $\dis \int_{z_0} \beta(z)^2 a(z) \,dz=\int^{z_1} \beta(z)^2 a(z) \,dz= +\infty$.

\end{enumerate}
\end{theorem}

A typical situation where these results apply is when $\dis\int^{z_1}\left(\frac1{b(z)}+d(z)+a(z)\right)dz=+\infty$  but $\dis\int_{z_0}\left(\frac1{b(z)}+d(z)+a(z)\right)dz$ $<+\infty$. Then $C_0^{\infty}(\O)$ is not dense in $\E$, but the only non trivial linear forms continuous on $\E$, vanishing on $C_0^{\infty}(\Omega)$, are supported on $\{z_0\}\times \M$. This means that a boundary condition must be chosen at $z=z_0$, but not at $z=z_1$.

Moreover if we have, for example, $\displaystyle{ \int^{z_1} a(z) \,dz=+\infty}$, the selfadjoint extension  $\tilde{A}$, defined from $A$ with an appropriate  boundary condition at $z=z_0$, will be unique. In particular, considering null Dirichlet boundary condition, $\tilde{A}$ will be the selfadjoint extension of $A$  constructed from the restriction of the bilinear form to $\E^0$.

\end{rem}

\section{Well-posedness of the Cauchy problem }\lab{S4}
Let $A$ and $\O$ be as in the previous section. We assume $A$ to be at least q.e.s.a. but not necessarily e.s.a.; we denote in the same way its unique selfadjoint extension with finite energy. We take functions $f\in \E$  and $g\in H$
 and consider the Cauchy problem
\beqn \lab{P}
(P) \left\{\begin{array}{cc}
 \pd {tt}\varphi + A\varphi &=0,\\
 \varphi(0,\cdot)&=f,\\
  \pd t \varphi (0,\cdot)&= g.
\end{array}\right.
\eeqn

\begin{theorem}\label{solution}

Under the hypotheses above, the problem (P) has a unique solution
$$
\phi \in C([0,\infty) ; \E)\cap C^1([0,\infty) ; H),
$$
and there exists a constant $C>0$ such that

$$
\forall \ t>0\ \ \ \|\phi(t,\cdot)\|_{\E}+ \|\pd t
\phi(t,\cdot)\|_{H}\leq C(\|f \|_{\E}+ \|g\|_{H}).
$$
 In this case, the energy
$$
E(\phi,t)=\frac{1}{2}\int_{\Omega} \left(a(z)\,(\pd t
\phi)^2+b(z)\,(\pd z \phi)^2+ c(z)\,|\nabla \phi|^2 + d(z) |\phi|^2
\right)d\mu
 $$
 is well-defined and conserved:
$$
\forall\  t>0 \ \ \ E(\phi, t)=\frac12 \left( \|g\|_H^2+ b(f,f)
\right).
$$

\end{theorem}

\noindent{\bf Proof: }

 Let $D$ be the domain  defined in \eq{dominioF},  given $f \in D$  and $g\in \E$, the solution of (P) is
given by (see, for example, \cite{W} and references therein)
\beq\label{onda}
\phi(t,\cdot)= \cos\Bigl(tA^{\frac12}\Bigr)f+
A^{-\frac12}\sin\Bigl(tA^{\frac12}\Bigr)g.
\eeq
Taking into account that $D(A^{\frac12})= \E$, we have
$\phi(t,\cdot) \in D$ and $\pd t\phi(t,\cdot)  \in \E$. That
$\phi(t,\cdot) $ and $\pd t\phi(t,\cdot) $ are continuous
vector-valued functions (in $D$ and in $\E$ respectively) rely on
a classical density argument we only sketch. For $\varepsilon >0$
we set $f_{\varepsilon}=(I+\varepsilon A)^{-1}f$,
$g_{\varepsilon}=(I+\varepsilon A)^{-1}g$ and
$\phi_{\varepsilon}=(I+\varepsilon A)^{-1}\phi$. Then $\pd
t\phi_{\varepsilon}(t,\cdot) \in D$ and $\pd
{tt}\phi_{\varepsilon}(t,\cdot) \in \E$, with their norms
uniformly bounded in $t$, while  $\phi_{\varepsilon}(t,\cdot)
\rightarrow \phi(t, \cdot) $ in $ D$  and $\pd
t\phi_{\varepsilon}(t,\cdot)\rightarrow \pd t\phi(t,\cdot)$ in
$\E$ when $\varepsilon \to 0$. The conclusion readily follows.

\medskip
 When $f \in \E$ and $g\in H$, we define $\phi(t,\cdot)$
by (\ref{onda}). Then
 $\phi(t,\cdot) \in \E$ and $\pd
t\phi(t,\cdot)  \in H$. The continuity results are obtained by
density arguments in the same way as above.

 The reader should notice that in this case we have
$\pd{tt}\phi(t,\cdot)+A(\phi(t,\cdot))=0$ in $\E'$, where $\E'$ is
the dual space of $\E$; hence $\phi$ is a weak solution of (P).
 Regarding the conservation of the energy, although the argument here is standard, we recall it for
its convenience. We assume first that $f \in D$ and
$g\in \E$. Then $\phi(t,\cdot)$ is a strong solution of (P) and we
have
\beq\label{yanose}
\int_{t_1}^{t_2}\int_{\Omega}a(z)\, \pd t\phi \, (\pd {tt} \phi + A
\phi)\, dt\, d\mu = 0.
\eeq
We consider each term separately, obtaining for the first one
\begin{eqnarray}  \label{tiempo}
\int_{\Omega}\int_{t_1}^{t_2}a(z)\, \pd t\phi\ \pd {tt} \phi \, dt\,
d\mu=\left.\frac12\int_{\Omega}a(z)\, ( \pd {t} \phi )^2 \,
d\mu\right|_{t_1}^{t_2},
\end{eqnarray}
and for the second one (see for instance \cite{K})
\begin{eqnarray}  \label{comilla}\fl
\int_{t_1}^{t_2}\int_{\Omega}\pd t\phi\ A \phi\, a(z)\, dt\, d\mu &=&
\int_{t_1}^{t_2}<\pd t\phi , A \phi >_H\, dt\nn
&=& \int_{t_1}^{t_2}b(\phi, \pd t\phi )\, dt\nn &=&
\frac12\left.\int_{\Omega} \left(a(z)\,(\pd t
\phi)^2+b(z)\,(\pd z \phi)^2+ c(z)\,|\nabla \phi|^2 + d(z) |\phi|^2
\right)d\mu  \ \right|_{t1}^{t_2}.\nn
\end{eqnarray}

Now, by (\ref{yanose}), adding (\ref{tiempo}) and (\ref{comilla}),
we have for all $t>0$
\begin{eqnarray}
  E(\phi,t)&=&\frac{1}{2}\int_{\Omega} \left(a(z)\,(\pd t
\phi)^2+b(z)\,(\pd z \phi)^2+ c(z)\,|\nabla \phi|^2 + d(z) |\phi|^2
\right)d\mu \nn &=&\frac12 \left( \|g\|_H^2+ b(f,f)
\right)\nonumber.
\end{eqnarray}

Again, by a density argument as before, this
result remains true when $f\in\E$ and $g\in H$.\cuad

\section{Propagation  of classical scalar fields  in static spherically symmetric spacetimes }\label{S5}


We consider a $(n+2)$-dimensional static and spherically symmetric \st\ with $n\geq 1$ and metric signature $(-+\dots+)$.
Due to the required isometries the more general line element can be written as
\beq \lab{met}
ds^2=-F(r)\,
dt^2+G(r)\,{dr^2} + r^2\ d\ell^2_{S^n}\,,  \eeq
where $d\ell^2_{S^n}$ is the metric on the unit $n$-sphere $S^n$ and $r$ in  $(0,+\infty)$. For a non-degenerate Lorentzian metric  $g_{ab}$, \eq{met} makes sense only for those values of $r$ such that $0<F(r)G(r)<+\infty$. On the other hand, since $g_{ab}(\pd t)^a(\pd t)^b=-F$, the Killing vector field $\pd t$ is timelike only in the region $F(r)>0$, and  so \st\, is static only in this region. Therefore, without loss of generality, from now on we shall restrict ourselves to the region where  $F(r)$ and $G(r)$ are both finite and positive. In addition  we shall assume that $F$ and $G$ are such that the condition $0<F(r),G(r)<+\infty$ holds in a finite  union of disjoint non empty  open subintervals   $(r^-_i,r^+_i)$  of $(0,+\infty)$ and $ F, F', G \in C^1(\dis \bigcup_{i=1}^m (r^-_i,r^+_i))$. If the spacetime is {\em asymptotically flat}, in the outer region $(r^-_m,+\infty)$, we can find coordinates such that $\dis \lim_{r\to+\infty} F(r)=\lim_{r\to+\infty} G(r)=1$.

Due to the required symmetries the more general energy-momentum tensor can be written as
\beq \lab{Tab} T^b_a={\rm diag}\{-\rho(r),p_r(r),p_\theta(r),\dots,p_\theta(r)\}\,,\eeq
where $\rho(r)$ is the energy density, and $p_r(r)$,  $p_\theta(r)$ are the principal pressures.
We shall assume that  $\rho(r)$ is bounded and the {\em dominant energy condition}\footnote {See for example \cite{Haw}} is  satisfied, which, in this case,  is equivalent to
\beq \lab{dec} |p_r(r)|,|p_\theta(r)|\leq\rho(r)<+\infty\,.\eeq

From \eq{met} and \eq{Tab} we get that Einstein's equations, i.e., $G_{ab}+\Lambda\,g_{ab}=8\pi T_{ab}$, become
\beq \lab{Gtt} G_t^t=-\frac n{2\,r^2}\,\left((n-1)\left(1-\frac1{G(r)}\right)+r\,\frac{G'(r)}{G(r)^2}\right)=-8\pi\,\rho(r)-\Lambda\,,\eeq
\beq  \lab{Grr}G_r^r=\frac n{2\,r^{2}}\,\left(\frac{r \,F'(r)}{F(r)G(r)}+(n-1)\left(\frac1{G(r)}-1\right)\right)=8\pi\,p_r(r)-\Lambda\,,\eeq
\beqa\lab{Gxx}
G_\theta^\theta&=\frac{F''(r)}{2 F(r) G(r)}-\frac{F'(r) G'(r)}{4 F(r)
G(r)^2}+\frac{(n-1) F'(r)}{2 r F(r)
G(r)}-\frac{F'(r)^2}{4 F(r)^2 G(r)}
\nn&-\frac{(n-1) G'(r)}{2r G(r)^2}-\frac{(n-2)
(n-1)}{2 r^2}\left(1-\frac1{G(r)}\right)
\nn&=8\pi\,p_\theta(r)-\Lambda\,,
\eeqa
where $\Lambda$  is the cosmological constant.
Furthermore, the local energy-momentum  conservation ($\cd a T^{ab}=0$) gives
\beq\lab{daTab} p_r'(r)= -\frac{\rho(r)+p_r(r)}{2}\,\frac{F'(r)}{F(r)}-n\ \frac{(p_r(r)-p_\theta(r))}{r}\,.\eeq
Of course, due to Bianchi's identities, \eq{Gtt}-\eq{daTab} are not independent. These are a system of three linear independent {\em ODE}\,'s and, in order to find the five unknown functions $F(r)$, $G(r)$, $\rho(r)$, $p_r(r)$ and $p_\theta(r)$, we have to provide  {\em equations of state} relating the functions $\rho(r)$, $p_r(r)$ and $p_\theta(r)$.

 From \eq{Gtt} and \eq{Grr} we can write down a  more handleable set of two equivalent  independent equations
\beq \lab{1} \left(r^{n-1}\left(1-\frac1{G(r)}\right)\right)'=\frac{2r^n} n\, \bigl(8\pi\,\rho(r)+\Lambda\bigr)\,,\eeq
\beq \lab{2} \ln'\Bigl(F(r)\,G(r)\Bigr)=\frac{16\pi} n\, \Bigl(\rho(r)+p_r(r)\Bigr)\,r\,G(r)\,,\eeq
which in the vacuum cases, leads readily to the solution.

Indeed,  if we for instance set $ \rho(r)=-p_r(r)=p_\theta(r)$, from \eq{daTab}  we immediately get that
\beqn p_r(r)=-\rho(r)=-\frac{C_1}{r^{2n}}\,,\eeqn
where the constant $C_1$ must be positive by \eq{dec}. Then, we find from \eq{1} that
\beqn \frac1{G(r)}=1-\frac{C_2}{r^{n-1}}+\frac{16\pi\,\,C_1}{ n(n-1)\,r^{2n-2}}-\frac{2\,\Lambda\,r^2}{n(n+1)}\,,\eeqn
where $C_2$ is a new arbitrary constant. And \eq{2} immediately gives $F(r)\,G(r)=C_3$, and we can always set the constant  $C_3=1$ by scaling the time. This family of solutions, depending on three parameters, includes the higher-dimensional generalization of Schwarzschild, de Sitter and Reissner-Nordstr\"{o}m geometries.

For future use, we shall prove the following result.

\begin{lem}\label{FG}
If\  \, $0<F(r),G(r)<+\infty$ in some interval $(r^-_i,r^+_i)$, then
\begin{enumerate}
\item   $F(r)G(r)$ is a nondecreasing function of $r$  in $(r^-_i,r^+_i)$, and then bounded in a neighborhood of $r^-_i$.

\item In the outer region of an asymptotically flat spacetime,  $F(r)G(r)$ is bounded.
\end{enumerate}

\end{lem}

{\bf Proof:}

{\it (i)}
As a consequence of the dominant energy condition \eq{dec}  the right hand side of \eq{2} cannot be negative, then $F(r)G(r)$ cannot be decreasing.

{\it (ii)} Since  $F(r)G(r)$ is nondecreasing,   we  get that $0<F(r)\,G(r)\leq1$ since $\dis \lim_{r\to+\infty} F(r)=\lim_{r\to+\infty} G(r)=1$.\cuad

 In these spacetimes, we shall consider the propagation of a
scalar field $\psi$ with Lagrangian density \beqn\label{lag}
\lag= -\frac{1}{2}\nabla^a\psi\, \cd a \psi -\frac{m^2}{2}\,\psi^2  , \eeqn
where  the constant $m$ is the mass of the field and $\nabla $  denotes the covariant
derivative (Levi-Civita connection).

 As usual, we obtain the field equations by requiring
that the action
$$ S=\int\lag(\cd a\psi,\psi,g_{ab})\sqrt{|g|}\ dt
d\mu\ $$
 be stationary under arbitrary variations of the fields $\delta
\psi$ in the interior of any compact region, but vanishing at its
boundary. Thus, we have the Euler-Lagrange equation
$$ \cd a\left(\dlag{ \cd a \psi}\
\right)=\dlag{ \psi}\ ,
$$
which, in our case,  becomes the Klein-Gordon equation
\beq\label{wave}
\cd a \nabla^a \psi\ = \Box \psi=\frac{\pd a \left(\sqrt{|g|}\ g^{ab}\,\pd b
\psi \right)} {\sqrt{|g|}}={m^2}\,\psi\,. \eeq
Therefore, we get from \eq{met} and \eq{wave} that the field equation may be written as
\beqn \pd {tt}
\psi =-\,A\psi\eeqn
where
\beqa \lab{A}\fl
A\psi =-\frac{1}{r^n}\sqrt{\frac{F(r)}{G(r)}}\left(\rule{0pt}{26pt}\pd r\left(r^n \sqrt{\frac{F(r)}{G(r)}}\, \pd r \psi \right)+ r^{n-2}\ \sqrt{{F(r)}{G(r)}}\ \Delta_{S^n}\psi \right)+{m^2}\,F(r)\,\psi\ ,\nn\eeqa
where $\Delta_{S^n}$ is the Laplacian on the unit $n$-sphere.
Then, by comparing with the operator defined in \eq{operador}, we get the identification of the coefficients
\beqa \lab{coef}
&a(r)= r^n \sqrt{\frac{G(r)}{F(r)}}\hs {.3},\hs {2}  &b(r)= r^n \sqrt{\frac{F(r)}{G(r)}}\hs {.3},\nn \hs {.3} &c(r)= r^{n-2} \sqrt{{F(r)}{G(r)}}\hs {.3},   &d(r)=m^2\,r^{n} \sqrt{{F(r)}{G(r)}}\ .\eeqa

\begin{rem}\rm
From \eq{met} we get that radial null geodesics satisfy $\dis\frac{dt}{dr}= \pm \sqrt{\frac{G(r)}{F(r)}}$. Then, if  $r_0$ and $r$ belong to the closure of a connected region where $0<F(s),G(s)<+\infty$, we find from \eq{coef}  that the coordinate time $t$ a radial photon takes to travel from $r$ to $r_0$ is
\beq\lab{t} T(r\to r_0)=\left| \int_{r}^{r_0} \sqrt{\frac{G(s)}{F(s)}}\ ds \right|=
\left| \int_{r}^{r_0} \sqrt{\frac{a(s)}{b(s)}}\ ds \right|\,.\eeq

We shall see that it is actually this time which plays a crucial role in the analysis of {\it e.s.a.}
when there is a horizon at $r_0$ ($r_0=r_i^+$ or $r_0=r_i^-$) in the spacetime, i.e., $T(r\to r_0)=+\infty$.
\end{rem}

\begin{lem}\lab{af}
In the outer region of an asymptotically flat spacetime one has $\dis \int^{+\infty} {a(r)}\, dr=+\infty$.
\end{lem}
\noindent{\bf Proof.}
If  $\dis \lim_{r\to+\infty} F(r)=\lim_{r\to+\infty} G(r)=1$ by \eq{coef} we have that $\dis \lim_{r\to+\infty} \frac{a(r)}{r^n}=1$, and then
$\dis \int^{+\infty} {a(r)}\, dr=+\infty$.\cuad

\begin{lem}\lab{5.2}

If $0<F(r),G(r)<+\infty$ in $(r^-_i,r^+_i)$, with $r_i^->0$, the three following statements are equivalent
\beqn\fl
 \,\int_{r^-_i} \frac{1}{b(r)}\, dr=+\infty,\hs {.5}
  \,\int_{r^-_i} {a(r)}\, dr=+\infty\hs {.5}  \text{and} \hs{.5}
\,\int_{r^-_i} \sqrt{\frac{a(r)}{b(r)}}\, dr=+\infty\,.
\eeqn
On the other hand, if $r^+_i$ is finite, the three following statements are equivalent
\beqn\fl
 \,\int^{r^+_i} \frac{1}{b(r)}\, dr=+\infty,\hs {.5}
  \,\int^{r^+_i} {a(r)}\, dr=+\infty\hs {.5}  \text{and} \hs{.5}
\,\int^{r^+_i} \sqrt{\frac{a(r)}{b(r)}}\, dr=+\infty\,.
\eeqn

\end{lem}

\noindent{\bf Proof.}

By \eq{coef} we have that $a(r)b(r)=r^{2n}$.
For $r_*<r<{r^*}<+\infty$,  we readily get  the inequalities
\beqn \frac{r_*^{2n}}{b(r)}<a(r)<\frac{{r^*}^{2n}}{b(r)}\hs 1\text{and}\hs 1
r_*^{n}\,\sqrt{\frac{a(r)}{b(r)}}<a(r)<{r^*}^{n}\,\sqrt{\frac{a(r)}{b(r)}}\,.\eeqn
Now, by integrating these expressions between $r_*$ and ${r^*}$, we get the result.\cuad

Observe that by the properties of the functions $F$ and $G$, under the hypotheses of lemma \ref{5.2} we have
\begin{itemize}
 \item $a$, $b$, $c$, $d \in  C^1\left(\rule{0pt}{10.5pt} (r^-_i,r^+_i)\right)$
 \item $a$, $b$, $c>0$ and $d\geq 0$ in $(r^-_i,r^+_i)$
 \item $a^{-1}$, $b^{-1}$, $c^{-1}  \in L_{loc}^1\left(\rule{0pt}{10.5pt}(r^-_i,r^+_i)\right).$
\end{itemize}

 Then, if we consider the operator defined by \eq{A} in  $\Omega=(r^-_i,r^+_i)\times S^n$, we have:

\begin{theorem}\lab{horizon}\
For $0<r^-_m<\infty$, let  $A$ be the  operator corresponding to the propagation of a scalar field in $\Omega=(r^-_m,\infty)\times S^n$
in a static, spherically symmetric and asymptotically flat spacetime where the dominant energy condition holds.  The three following statements are equivalent:
 \begin{enumerate}

\item The time
$\ T(r\to r^-_m)$ is infinite.

\item
$A$ is a q.e.s.a. operator.

\item $A$ is an e.s.a. operator.

\end{enumerate}
\end{theorem}

Or, in other words, $A$ is e.s.a. if and only if  a radial photon needs an infinite amount of time to get $r^-_m$.

\noindent{\bf Proof:}\

{\it (i)} $\Rightarrow$ {\it (ii)} and {\it (iii)}: By lemma \ref{af} we have that $\dis \int^{+\infty} {{a}(r)}\, dr=+\infty$.
On the other hand, if $\ T(r\to r^-_m)=+\infty $, it follows by \eq{t} that $\dis \int_{r^-_m} \sqrt{\frac{a(r)}{b(r)}}\, dr=+\infty$,
and then from lemma \ref{5.2} we have $\dis \int_{r^-_m} {a(z)}\, dz= +\infty$. Therefore, it follows from theorem \ref{qesa1}  that the operator $A$ is {\it q.e.s.a} and from theorem \ref{coresa} {\it(i)}  that the operator $A$ is {\it e.s.a}.

{\it (ii)} $\Rightarrow$ {\it (i)}: Conversely, assume that $\ T(r\to r^-_m)<+\infty $, then $\dis \int_{r^-_m} \sqrt{\frac{a(r)}{b(r)}}\, dr<+\infty$. And it immediately follows from lemma \ref{5.2} that $\dis  \int_{r^-_m} {a(r)}\, dr<+\infty$ and $\dis \int_{r^-_m} \frac 1{b(r)}\, dr<+\infty$. On the other hand,  since $F(r)G(r)$ is bounded by lemma \ref{FG}, $\dis \int_{r^-_m} {{d}(r)}\, dr= m^2\,\int_{r^-_m} r^{n} \sqrt{{F(r)}{G(r)}}\, dr<+\infty$. Therefore, it follows from theorem \ref{qesa1}  that the operator $A$ is not {\it q.e.s.a}.

{\it (iii)} $\Rightarrow$ {\it (ii)}: This is obvious by definition.\cuad

\begin{rem}\rm
Note that  the boundedness of $F(r)G(r)$ is only used in the proof of the sufficiency of the condition
$\ T(r\to r^-_m)=+\infty $, to guarantee that $d(r)$ is integrable at $r^-_m$. Therefore, for massless fields, since in  this case $d(r)\equiv0$  the theorem follows without invoking  any energy condition.

Similar results also follow from remark \ref{01}  and lemma \ref{5.2} at internal horizons.

\end{rem}

\section{Examples}\lab{S6}

\subsection{$(n+2)$-dimensional punctured Minkowski \st}\lab{4.3}

Here we consider the flat $(n+2)$-dimensional  Minkowski \st\ with a removed spatial point. We chose the origin of coordinates at this point and then the line element can be written as
\beqn
ds^2=-dt^2+{dr^2} + r^2\ dl^2_{S^n}\ ,\eeqn
where $-\infty<t<+\infty$ and $ 0<r<+\infty$.
This \st\ has a time-like singular boundary along the $t$ axis.
In this case, $\Omega=(0,\infty)\times S^n$ and  $F(r)=G(r)=1$, so  the coefficients in \eq{coef} are $a(r)=  b(r)={r^n}$, $ c(r)= r^{n-2}$ and $d(r)=m^2\,r^n$.   The operator $A$ in \eq{A} turns out to be
\beqan A\psi =-\frac 1{r^n}\ \pd r\left(r^n \, \pd r \psi \right)-\frac{1}{r^2}\,
\Delta_{S^n} \psi+m^2\,\psi\ ,\eeqan
which formally is nothing but $-\Delta + m^2$.

Now,  for $n\geq 1$, we have that
$\dis \int^{+\infty} a(r)\, dr=+\infty$  and
$\dis
\int_0 \frac{dr}{b(r)} =+\infty\,.$
Then it immediately follows from theorem \ref{q.e.s.a.} that $A$ is a {\it q.e.s.a.} operator for every $m^2\geq0$ and every $n\geq1$.

We turn now  to explore whether $A$ is an {\it e.s.a.} operator too. Taking into account that
$ d(z)/{a(z)}=m^2$,  $\dis \int_{0} a(z)\, dz= \int_{0} r^n\, dz<+\infty$ and $\dis  \int^{+\infty}a(z)\, dz=+\infty$,
we can apply corollary \ref{coro}.

 Now, for $0<r_1<+\infty$,  we have
\beqan \beta_0(r)=\int_r^{r_1} \frac{du}{b(u)}= \cases{
-\ln\left(\frac{r}{{r_1}}\right) & \quad
{ if } $n=1$ \\
\frac{r^{1-n}-{r_1}^{1-n}}{n-1} &\quad
  \text{ if } $n\geq 2$ }\ .\eeqan
Thus,
$$\int_0^{r_1} \beta_0^{\,2}(r) a(r) dr<+\infty\ $$ if and only if $n=1,2$.
 Therefore, it immediately  follows from corollary \ref{coro} that $A$ is  an {\it e.s.a.} operator  only if $n\geq3$. This is a well known result, see for instance \cite{BS,RS}.

\subsection{$(n+2)$-dimensional anti-Schwarzschild ($M<0$) \st}
Here we consider the $(n+2)$-dimensional  \st\ with line element
\beqan
ds^2=-\left(1+\frac{r_s^{n-1}}{r^{n-1}}\right)dt^2+{ \left(1+\frac{r_s^{n-1}}{r^{n-1}}\right)^{-1}}{dr^2} + r^2\ d\Omega^2_{S^n}\ ,\eeqan
where  $-\infty<t<+\infty$, $0<r<+\infty $, $r_s$ is a positive constant and $n\geq2$\footnote{The case $n=1$ is 3-dimensional Minkowski \st\ already discussed in \ref{4.3}}. This spacetime has a naked timelike singularity at $r=0$ where some components of the Weyl  tensor diverge.

In this case, $\Omega=(0,\infty)\times S^n$ and we get from \eq{coef} that the  coefficients of the operator $A$ are
\beqn \fl
a(r)= \frac{r^{2n-1}}{ r^{n-1}+r_s^{n-1}} \hs{.3},\hs{.3} b(r)={r}( r^{n-1}+r_s^{n-1})\hs{.3},\hs{.3}c(r)= r^{n-2}\hs{.3}\text{and} \hs{.3}d(r)=m^2\,r^n \ .\eeqn

We get therefore
\beqn \int_0\frac{dr}{b(r)} = +\infty\  \ \mbox{and} \ \  \int^{+\infty} a(r)\, dr=+\infty .\eeqn
Then it immediately follows from theorem \ref{q.e.s.a.} that $A$ is a {\it q.e.s.a.} operator for every $m^2\geq0$ and every $n\geq2$.

For $m=0$ and $n=2$, we have already proved in \cite{GST} that $A$ is not an {\it e.s.a.} operator. Here, we shall analyze the general case.

We first consider the case $m=0$. Taking into account that
\beqn \int_0a(r)\,dr<+\infty\, , \ \ \int^{+\infty} a(r)\, dr=+\infty\ \mbox{and}\ \  d(z)=0, \eeqn we can apply corollary \ref{coro}.

For $0<r<r_s$,  we have
\beqn \beta_0(r)=\int_r^{r_s} \frac{ds}{b(s)} =\frac{-1}{ r_s^{n-1}(n-1)}\ln\left(\frac{2\,r^{n-1}}{r^{n-1}+r_s^{n-1}}\right)\ ,
\eeqn
and
$$\lab{bc} \int_0^{r_s} \beta_0^{\,2}(r) a(r) dr<+\infty\,.$$
Thus, in the massless case, $A$ is not an {\it e.s.a.} operator for every $n\geq2$ thanks to the corollary \ref{coro}.

For  $m^2>0$ we cannot apply corollary \ref{coro} since $ d(z)/{a(z)}$ is not bounded near $0$. Nevertheless, the ordinary differential equation  \eq{ODE}, satisfied by the function $\alpha(z)$ of lemma \ref{lIII}, becomes in this case
\beqn - \Bigl(r\,( r^{n-1}+r_s^{n-1})\,\alpha'(r)\Bigr)'+m^2\,r^n\, \alpha(r)=0\ ,\eeqn
and a straightforward computation shows  that
\beqn  \alpha(z)=\alpha(0)\Biggl(1+\frac{m^2\,r_s^2}{(n+1)^2}\,\left(\frac{r}{r_s}\right)^{n+1}
\hs{-.3} -\frac{m^2\,r_s^2}{2n(n+1)}\,\left(\frac{r}{r_s}\right)^{2n}\ +\dots\Biggr) \,\eeqn
 near $0$.
Furthermore, since by lemma \ref{lIII} $\alpha(r)$ is positive and increasing in $(0,r_s)$, and by definition $\alpha(r_s)=1$, we get that $0<\alpha(0)<1$.

Therefore
\beqn \beta(r)=\alpha(r)\,\int_r^{r_s} \frac{ds}{b(s)\alpha(s)^2} <\frac{1}{ \alpha(r)} \,\int_r^{r_s} \frac{ds}{b(s)}<\frac{1}{ \alpha(0)}\,\beta_0(r)\,
\eeqn
and
\beqn  \int_0^{r_s} \beta^{\,2}(r) a(r) dr<\frac{1}{ \alpha(0)^2}\,\int_0^{r_s} \beta_0^{\,2}(r) a(r) dr<+\infty\,.\eeqn
It follows from theorem \ref{esa} {\it(ii)} that  $A$ is not an {\it e.s.a.} operator for every $n\geq2$ and $m^2\geq0$.

\begin{rem}
Note that the estimate
\beqn \beta(z)=\alpha(z)\,\int_z^{1} \frac{ds}{b(s)\alpha(s)^2} <\frac{1}{ \alpha(z)} \,\beta_0(z)\,, \eeqn
when $\alpha(0)\neq0$, also gives a necessary and sufficient condition for {\it e.s.a.} in terms of $\beta_0(z)$ only.

For analytic $b(z)$ and  $d(z)$, as in our example, $\alpha(0)\neq0$ if  one of the roots of the indicial polynomial of \eq{ODE} is zero and the other non positive, which requires that
\beqn \lim_{z\to0^+} \frac{z^2d(z)}{b(z)}=0\hs 1 \text{and}\hs 1 \lim_{z\to0^+} \frac{z\,b'(z)}{b(z)}\geq1\,.\eeqn

\end{rem}

\subsection{$(n+2)$-dimensional Schwarzschild-Tangherlini  \st}

Here we consider the $(n+2)$-dimensional  \st\ with line element
\beqn
ds^2=-\left(1-\frac{r_s^{n-1}}{r^{n-1}}\right)dt^2+{ \left(1-\frac{r_s^{n-1}}{r^{n-1}}\right)^{-1}}{dr^2} + r^2\ d\Omega^2_{S^n}\ ,\eeqn
where $r_s$ is a positive constant, $-\infty<t<+\infty$, $0<r<r_s$ or $r_s<r<+\infty$ and $n\geq2$. This spacetime has a  spacelike irremovable singularity at $r=0$ where some components of the Riemann tensor diverge and an event horizon at $r=r_s$, the latter may be removed by introducing suitable coordinates and extending the manifold to obtain a maximal analytic extension \cite{KRU}. As already mentioned, our wave formulation only makes sense in the static region ($r_s<r<+\infty$), and we will use it to explore the properties of the wave equation \eq{wave} in this region.

Thus, we consider the operator $A$ given by \eq{A} in   $\Omega=(r_s,\infty)\times S^n$, and we see from \eq{coef} that
\beqn\fl
a(r)= \frac{r^{2n-1}}{ r^{n-1}-r_s^{n-1}} \hs{.5},\hs{.5} b(r)={r}( r^{n-1}-r_s^{n-1})\hs{.5}\text{and} \hs{.5}d(r)=m^2\, r^n \ .\eeqn
Now,
we get from \eq{t} that
\beqn
T(r\to r_s)=\int_{r_s}^r \left(\frac{a(s)}{b(s)}\right)^{\frac{1}{2}}\ ds= \int_{r_s}^r \frac{s^{n-1}}{s^{n-1}-r_s^{n-1}}\ ds =+\infty \ . \eeqn
Therefore, it immediately follows from theorem \ref{horizon} that $A$ is an {\it e.s.a.} operator in $\Omega=(r_s,\infty)\times S^n$ for every $n\geq2$ and any $m^2\geq0$, and the Cauchy problem is well-posed without requiring any boundary condition at the event horizon.

\subsection{$(n+2)$-dimensional Reissner-Nordstr\"{o}m \st}

Here we consider the $(n+2)$-dimensional  \st\ with line element
\beqn\fl
ds^2=-\left(1-\frac{r_s^{n-1}}{r^{n-1}}+ \frac{q^{2n-2}}{4\,r^{2n-2}}\right)dt^2+{ \left(1-\frac{r_s^{n-1}}{r^{n-1}}+ \frac{q^{2n-2}}{4\,r^{2n-2}}\right)^{-1}}{dr^2} + r^2\ d\Omega^2_{S^n}\ ,\eeqn
where $r_s$ and $q^2$ are positive constants and $n\geq2$\footnote{The case $n=1$ is again 3-dimensional Minkowski \st\ already discussed in \ref{4.3}}. If $q^2>r_s^2$ the metric is non-singular everywhere except for the timelike irremovable repulsive singularity at $r=0$. If $q^2\leq r_s^2$, the metric also has singularities at $r_+$ and $r_-$, where $r_{\pm}^{n-1}=(r_s^{n-1}\pm\sqrt{r_s^{2n-2}-q^{2n-2}})/2$; it is regular in the regions defined by $\infty>r>r_+$, $r_+>r>r_-$ and $r_->r>0$ (if $q^2=r_s^2$ only the first and the third regions exist). As in the Schwarzschild case, these singularities may be removed by introducing suitable coordinates and extending the manifold to obtain a maximal analytic extension \cite{GB,C}. The first and the third regions are both static, whereas the second region (when it exists) is spatially homogeneous but not static.

We shall study the properties of the  wave equation  in the static regions. For convenience we shall analyze separately the three cases. Note that, in the three cases this spacetime is asymptotically flat.

\subsubsection{Case $q^2>r_s^2$} This \st\ has only a naked timelike irremovable repulsive singularity at $r=0$. In this case, we consider the operator $A$ given by \eq{A} in  $\Omega=(0,\infty)\times S^n$, and from \eq{coef} we have
\beqn\fl
a(r)= \frac{r^{n}}{\dis 1-\frac{r_s^{n-1}}{r^{n-1}}+ \frac{q^{2n-2}}{4\,r^{2n-2}}}\ ,\hs{.5}b(r)= r^n-r_s^{n-1}r+\frac{q^{2n-2}}{4\,r^{n-2}}\hs{.5}\text{and} \hs{.5}d(r)=m^2 r^n\ . \eeqn

Then
\beqn \int_{0}\frac{dr}{b(r)}+a(r)+d(r)\, dr < +\infty\,.\eeqn
Hence it follows from theorem \ref{q.e.s.a.} {\it(ii)}\ \ that $A$ is not even a {\it q.e.s.a.} operator in this case, for every $n\geq2$ and any $m^2\geq0$. Therefore, in contrast to the anti-Schwarzschild case, in order to have a well-possed Cauchy problem a boundary  condition at the singularity must be given.

\subsubsection{Case $r_s^2=q^2$ (extreme case)}
This \st\  also has a removable singularity at $r_*= 2^{\frac{-1}{n-1}}r_s$. In this case, we consider the operator $A$ given by \eq{A} in two regions $\Omega=(0,r_*)\times S^n$ or $\Omega=(r_*,\infty)\times S^n$.

We get from \eq{coef} that
\beqn \fl
a(r)=\frac{r^{3n-2}}{\left(r^{n-1}- r_*^{n-1}\right)^2}\hs{.5},\hs{.5} b(r)=  \frac{\left(r^{n-1}- r_*^{n-1}\right)^2}{r^{n-2}}\hs{.5}\text{and} \hs{.5}d(r)=m^2\, r^n \ .\eeqn

We first consider the outer region  ($r_*<r<+\infty$). In this case, we get from \eq{t} that
\beqn
T(r\to r_*)=\int_{r_*}^r \left(\frac{a(s)}{b(s)}\right)^{\frac{1}{2}}\ ds= \int_{r_*}^r \frac{s^{2n-2}}{\left(s^{n-1}- r_*^{n-1}\right)^2}\ ds =+\infty \ . \eeqn
Therefore, it follows from theorem \ref{horizon} that $A$ is an {\it e.s.a.} operator in $\Omega=(r_*,\infty)\times S^n$ for every $n\geq2$ and any $m^2\geq0$, and the Cauchy problem is well-posed without requiring any boundary condition at the event horizon.

Regarding the inner region  $0<r<r_*$, we get that
\beqn\int_{0}\left( \frac 1{b(z)}+d(z)+a(z)\right)\, dz <+ \infty\,.\eeqn
Hence it follows from theorem  \ref{qesa1} that $A$ is not even a {\it q.e.s.a.} operator, for every $n\geq2$ and any $m^2\geq0$.

However,  we have
\beqn
\int^{r_*}a(r)\,dr =\int^{r_*}\frac{r^{3n-2}}{\left(r^{n-1}- r_*^{n-1}\right)^2}\,dr= +\infty\,,\eeqn
so it follows from  remark  \ref{01}  that in order to have a well-posed Cauchy problem in $\Omega=(0,r_*)\times S^n$ a boundary  condition at the singularity ($r=0$) must be given but not at the horizon ($r=r_*$).

\subsubsection{Case $r_s^2>q^2$}
This \st\ has, besides the timelike irremovable repulsive singularity at $r=0$, two removable  singularities at $r_+$ and $r_-$. In this case, we consider the operator $A$ given by \eq{A} in two regions $\Omega=(0,r_-)\times S^n$ or $\Omega=(r_+,\infty)\times S^n$, by abuse of notation we call $A$ these two different operators.

From \eq{coef} we can write
\beqan \fl
a(r)=\frac{r^{3n-2}}{\left(r^{n-1}- r_-^{n-1}\right)\left(r^{n-1}- r_+^{n-1}\right)}\ ,\hs 1  b(r)=  \frac{\left(r^{n-1}- r_-^{n-1}\right)\left(r^{n-1}- r_+^{n-1}\right)}{r^{n-2}}\\ \hs 2\text{and}\hs 1\,d(r)=m^2\, r^n \ .\eeqan

We first consider the outer region  ($r_+<r<+\infty$). In this case, we get from \eq{t} that
\beqn\fl
T(r\to r_*)=\int_{r_*}^r \left(\frac{a(s)}{b(s)}\right)^{\frac{1}{2}}\ ds
= \int_{r_+}^{r} \frac{s^{2n-2}}{\left(s^{n-1}- r_-^{n-1}\right)\left(s^{n-1}- r_+^{n-1}\right)}\ ds =+\infty \ . \eeqn
Therefore, it follows from theorem \ref{horizon} that $A$ is an {\it e.s.a.} operator in $\Omega=(r_+,\infty)\times S^n$ for every $n\geq2$ and any $m^2\geq0$, and the Cauchy problem is well-posed without requiring any boundary condition at the event horizon.

Regarding the inner region  $0<r<r_-$, we get
\beqn\int_{0}\left( \frac 1{b(z)}+d(z)+a(z)\right)\, dz <+ \infty\,.\eeqn
Hence it follows from theorem  \ref{qesa1} that $A$ is not even a {\it q.e.s.a.} operator, for every $n\geq2$ and any $m^2\geq0$.

However,  we have
\beqn \fl
\int^{r_*}a(r)\,dr =\int^{r_*}\frac{r^{3n-2}}{\left(r_-^{n-1}- r^{n-1}\right)\left(r_+^{n-1}- r^{n-1}\right)}\,dr= +\infty\,,\eeqn
so it follows from  remark  \ref{01}  that in order to have a well-posed Cauchy problem in $\Omega=(0,r_-)\times S^n$ a boundary  condition at the singularity ($r=0$) must be given but not at the horizon ($r=r_-$).

\end{document}